\documentclass[fleqn,usenatbib]{mnras}

\usepackage{newtxtext}
\usepackage{newtxmath}
\usepackage{xcolor}

\usepackage[T1]{fontenc}

\DeclareRobustCommand{\VAN}[3]{#2}
\let\VANthebibliography\thebibliography
\def\thebibliography{\DeclareRobustCommand{\VAN}[3]{##3}\VANthebibliography}

\usepackage{graphicx}	
\usepackage{amsmath}	
\usepackage{amssymb} 	

\graphicspath{{./}{figures/}}

\newcommand{\mf}{\textsf{m12f}}
\newcommand{\mi}{\textsf{m12i}}
\newcommand{\mm}{\textsf{m12m}}

\title[Shapes of MW-Mass Galaxies with SIDM]{Shapes of Milky-Way-Mass Galaxies with Self-Interacting Dark Matter}

\author[Vargya et al.]
{Drona Vargya,$^{1}$\thanks{E-mail: vad@sas.upenn.edu (DV)}
Robyn Sanderson,$^{1,2}$
Omid Sameie,$^{3}$
Michael Boylan-Kolchin,$^{3}$
Philip F. Hopkins,$^{4}$
\newauthor
Andrew Wetzel,$^{5}$
Andrew Graus$^{3}$
\\
$^{1}$Department of Physics and Astronomy, University of Pennsylvania, 209 South 33rd Street, Philadelphia, PA 19104, USA\\
$^{2}$Center for Computational Astrophysics, Flatiron Institute, 162 Fifth Avenue, New York, NY 10010, USA\\
$^{3}$Department of Astronomy, The University of Texas at Austin, 2515 Speedway, Stop C1400, Austin, TX 78712, USA\\
$^{4}$TAPIR, California Institute of Technology, MC 350-17, Pasadena, CA 91125, USA\\
$^{5}$Department of Physics and Astronomy, University of California, Davis, One Shields Avenue, Davis, CA 95616, USA\\
}

\date{Accepted XXX. Received YYY; in original form ZZZ}

\pubyear{2022}

\begin{document}
\label{firstpage}
\pagerange{\pageref{firstpage}--\pageref{lastpage}}
\maketitle

\begin{abstract}
Self-interacting dark matter (SIDM) models offer one way to reconcile inconsistencies between observations and predictions from collisionless cold dark matter (CDM) models on dwarf-galaxy scales.  In order to incorporate the effects of both baryonic and SIDM interactions, we study a suite of cosmological-baryonic simulations of Milky-Way (MW)-mass galaxies from the Feedback in Realistic Environments (FIRE-2) project where we vary the SIDM self-interaction cross-section $\sigma/m$.  We compare the shape of the main dark matter (DM) halo at redshift $z=0$ predicted by SIDM simulations (at $\sigma/m=0.1$, 1, and 10 cm$^2$ g$^{-1}$) with CDM simulations using the same initial conditions.  In the presence of baryonic feedback effects, we find that SIDM models do not produce the large differences in the inner structure of MW-mass galaxies predicted by SIDM-only models.  However, we do find that the radius where the shape of the total mass distribution begins to differ from that of the stellar mass distribution \emph{is} dependent on $\sigma/m$.  This transition could potentially be used to set limits on the SIDM cross-section in the MW.
\end{abstract}




\section{Introduction}
\label{sec:introduction}

The cold dark matter (CDM) plus dark energy $\left( \Lambda \text{CDM} \right)$ cosmological model has been the most successful model for understanding the large-scale structure of the Universe.  However, on length scales smaller than $\sim 1$ Mpc and masses smaller than $\sim 10^{11}$ M$_\odot$, there are challenges to this model from discrepancies between predictions from CDM-only simulations and observational data.  On dwarf galaxy ($M_\star \lesssim 10^9$ M$_{\odot}$) scales, these discrepancies include the core-cusp, diversity, missing satellites, too-big-to-fail (TBFT), and planes-of-satellites ``problems'' \citep[][and references therein]{2017ARA&A..55..343B, 2018PhR...730....1T}.  The \textit{core-cusp} problem \citep[][]{moore1994, kuzio2008, oh2008, walker2011, oh2015} arises from observational evidence that the cores of dark matter (DM) dominated galaxies are less dense and less cuspy (in inner density profile slope) than predicted by CDM-only simulations.  The \textit{diversity} problem, both in the field and among the Milky Way's (MW) satellites \citep[][]{oman2015,kaplinghat2019}, arises from observational evidence that galaxies' rotation curves appear to be more varied than in CDM simulations.  The \textit{missing satellites} problem \citep[][]{1999ApJ...522...82K, 1999ApJ...524L..19M} refers to the smaller number of observed satellite (dwarf) galaxies around the MW and the Local Group than the number of subhaloes predicted by CDM-only simulations.  The \textit{too-big-to-fail} problem \citep[][]{mbk2011, mbk2012, gk2014} arises from a mismatch in the central densities of satellite and field galaxies, which are smaller than predicted by CDM-only simulations;  it is expected that haloes so massive are resistant to star formation suppression from re-ionization (thus ``too big to fail'' in forming stars).  Finally, the \textit{planes-of-satellites} phenomenon refers to the apparent alignment of the orbital planes of satellite galaxies in the MW (e.g. \citealt{1976MNRAS.174..695L, 2005A&A...431..517K, 2012MNRAS.423.1109P, 2018A&A...619A.103F, 2020MNRAS.491.3042P}; though see also \citealt{2020MNRAS.494..983R}), M31 \citep[e.g.][]{2013Natur.493...62I, 2013ApJ...766..120C}, and Centaurus A \citep[][]{2018Sci...359..534M}, which is not commonly seen in CDM-only simulations \citep[e.g.][]{2009MNRAS.399..550L, 2014ApJ...784L...6I}; however, transient coplanar configurations are seen when baryons are included \citep[e.g.][]{2017MNRAS.466.3119A, 2019MNRAS.488.1166S, 2021MNRAS.tmp..945S}.

In order to solve these challenges on small scales without affecting large-scale structure, solutions within the CDM framework have been proposed to reconcile observations with predictions through a more complete incorporation of the baryonic physics \citep[][]{navarro1996, mashchenko2008, 2012ApJ...759L..42P, pontzen2012, governato2012, brooks2014, 2015MNRAS.454.2092O}.  Alternatively, the aforementioned discrepancies may hint toward a theory of DM beyond the CDM model.

Self-interacting dark matter (SIDM) models \citep[][]{2000PhRvL..84.3760S} assume DM particles can exchange energy and momentum via a force mediator with a cross-section close to the regime of the strong interaction \citep[][]{ahn2005, ackerman2009, feng2009, arkani2009, loeb2011, tulin2013}.  On galactic scales, the DM interaction rate becomes comparable to Hubble time-scale only deep inside of the gravitational potential, leaving the outer radii and extragalactic scales intact \citep[][]{2013MNRAS.430...81R, vogelsberger2016, sameie2019, 2020arXiv200606623B}.  In the inner regions of DM-dominated systems, energy-exchange through self-interactions will result in an isothermal density profile if the SIDM local collision rate $\Gamma_\mathrm{scatter} \propto \rho_\mathrm{DM} \, \sigma_\mathrm{x}/m_\mathrm{x} \geq 1$, where $\rho_\mathrm{DM}$ is the DM density and $\sigma_\mathrm{x}/m_\mathrm{x}$ is the self-interaction cross-section per unit mass.  The value of $\sigma_\mathrm{x}/m_\mathrm{x}$ is constrained by observations of galaxy clusters \citep[][]{yoshida2000a, randal2008, peter2013} and of the internal stellar kinematics of MW satellites \citep[][]{koda2011, zavala2013, valli2018, correa2020, hayashi2020}.

Cosmological DM-only simulations of SIDM models have made baseline predictions for their velocity profiles, density profiles, halo shapes, and substructures \citep[][]{2013MNRAS.430...81R, 2013MNRAS.430..105P, vogel2012}.  These simulations predict isothermal density profiles and spherical shapes for DM haloes and their substructures.  Introducing baryonic components in SIDM haloes couples the central DM densities to the baryonic potential \citep[][]{kaplinghat2014, elbert2018, 2018MNRAS.479..359S}, leading to substantial differences from DM-only predictions when baryons are dynamically important.  This suggests that a plausible explanation of the observed diversity in the DM distribution in field galaxies and the MW's satellites could be that it is a byproduct of baryonic mass assembly and DM self-interactions \citep[][]{creasey2017, kamada2017, ren2019, despali2019, sameie2020a, sameie2020b}.   

DM self-interactions also lead to more spherical halo shapes in SIDM than CDM \citep[][]{peter2013}.  Cosmological mass assembly in CDM creates triaxial DM haloes: since angular momentum exchange is inefficient, the DM particles retain substantial memory of their initial in-fall directions, resulting in haloes with ellipsoidal minor-to-major axis ratios as low as $c/a \sim 0.5$ \citep[][]{2011MNRAS.416.1377V, 2016MNRAS.462..663B}.  In pure SIDM, particles can more efficiently exchange angular momentum through interactions, isotropizing their orbits until $c/a \sim 1$.  However, if baryons dominate the gravitational potential, DM self-interactions tie the DM halo shapes to the baryonic distribution.  Semi-analytic modeling suggests that the SIDM density profile should scale with the \emph{total} gravitational potential \citep[][]{kaplinghat2014}.  If baryons dominate the central density of galaxies, the shape of the SIDM distribution should then follow that of the baryons.  $N$-body SIDM simulations of MW-mass systems with embedded baryonic discs support these predictions \citep[][]{2018MNRAS.479..359S}, as do the SIDM cosmological-baryonic simulations of slightly more massive disc galaxies (at $z \sim 0.2$) by \citet{despali2019}.

In this work and a companion paper \citep[][]{2021arXiv210212480S}, we examine high-resolution cosmological-baryonic simulations of MW-mass galaxies from the ``Feedback In Realistic Environments" (FIRE) project.  The initial conditions and CDM simulations are part of the second generation of simulations, the FIRE-2 suite \citep[][]{2018MNRAS.480..800H}; we also study the same initial conditions resimulated with several different SIDM cross-sections.  As in the original FIRE-2 suite, gravity and hydrodynamics are solved with \texttt{GIZMO} and baryonic feedback is implemented with the FIRE-2 feedback recipes, which are held constant across all runs \citep[for exact details, see][and Section \ref{sec:sims} of this paper]{2021arXiv210212480S}.  Simulations with SIDM use the Monte Carlo approach to scattering described in \citet{2013MNRAS.430...81R}.  We also resimulate some haloes without baryons, in both CDM and SIDM, to isolate feedback effects.  Our goal in this work is to gauge the extent to which halo shapes can serve as a discriminator between CDM and SIDM and the extent to which this depends on the self-consistent inclusion of baryonic physics. 

The CDM cosmological-baryonic versions of these simulations have previously been shown to produce a realistic population of satellite galaxies around MW-mass hosts that does not suffer from the \textit{missing satellites} or \textit{TBFT} problems of small-scale structure formation \citep[][]{2016ApJ...827L..23W, 2019MNRAS.487.1380G, 2020MNRAS.491.1471S, 2021MNRAS.tmp..945S}.  Furthermore, studies across mass scales have shown that the \textit{core-cusp} \citep[][]{2015MNRAS.454.2981C, 2015MNRAS.454.2092O, 2016ApJ...820..131E} and \textit{diversity} \citep[][]{2018MNRAS.473.1930E} problems are also mitigated with this feedback implementation.  Other groups find similar results with different physics implementations \citep[e.g.][]{2013ApJ...765...22B, 2014ApJ...786...87B, 2016MNRAS.456.3542T, 2016MNRAS.457L..74D}.  Since baryonic physics can thus at least partially reconcile observations with the standard $\Lambda$CDM cosmological model, we must also carefully gauge whether SIDM, combined with baryonic feedback, \emph{over}-corrects the potential small-scale problems for $\Lambda$CDM. 

This work is organized as follows.  In \S \ref{sec:methods}, we outline the method used to determine the shapes of haloes.  In \S \ref{sec:sims}, we detail the suite of simulations used in this study.  In \S \ref{sec:results}, we compare the results for the densities, velocities, scattering rates, shapes, and triaxiality between FIRE-2 MW-mass CDM and SIDM with previous results from simulations and observations.  In \S \ref{sec:discussion}, we discuss the results.  In \S \ref{sec:conclusion}, we give a summary of our results and conclusions.

\section{Methods of Determining Shapes}
\label{sec:methods}

To determine halo shapes, we use the iterative algorithm introduced in \citet{1991ApJ...378..496D} \citep[also see][]{2006MNRAS.367.1781A, 2011MNRAS.416.1377V, 2017ARA&A..55..343B, 2018MNRAS.479..359S, 2019MNRAS.488.3646R}.  This procedure fits a triaxial ellipsoid to the approximate isodensity surface of particles starting from a series of spherical radii $\{r\}$ from the galactic-centre by determining the weighted inertia tensor for particles inside each $r$.  We determine the axis ratios of these ellipsoids for each separate species in the simulations (DM, stars, and gas) and for the total mass distribution (which includes all particles from each species).  The axis lengths of the ellipsoids are labelled as $a(r) \ge b(r) \ge c(r)$, where $a(r)$, $b(r)$, and $c(r)$ are lengths of the major, intermediate, and minor semi-axes, respectively.  The axis lengths are defined as functions of $r$ to allow for changing shapes at different radii.  The axis ratios are then defined as:
\begin{equation}
	s(r) \equiv \frac{c(r)}{a(r)} \,, \quad p(r) \equiv \frac{c(r)}{b(r)} \,, \quad q(r) \equiv \frac{b(r)}{a(r)} \,.
\end{equation} 

We begin the iterative algorithm by calculating the weighted (or ``reduced'') inertia tensor, which is a symmetric matrix defined as
\begin{equation}
    \tilde{I}_{ij}(r) = \sum_{n=1}^{N_c} \frac{m_n \, x_{i,n} \, x_{j,n}}{d_n^2(r)} \bigg/ \sum_{n=1}^{N_c} m_n\,, \quad i,j \in \left\{ 1,2,3 \right\}\,,
\end{equation}
where $N_c$ is the number of particles within the ellipsoid of each component (or species), $m_n$ is the $n$th particle mass, and $x_{i,n}$ is the $i$th coordinate of the $n$th particle for each component with respect to a Cartesian coordinate system. In our final, best-fitting ellipsoid coordinate system, $x_{1}$ ($x_2$, $x_3$) corresponds to the distance along the major (intermediate, minor) axis. The tensor is ``reduced'' by normalizing the particle positions $\{x\}$ by the ellipsoidal distance $d_n(r)$\footnote{The unweighted inertia tensor $I_{ij}(r)$ (without the tilde), does not ``reduce'' the matrix with the ellipsoidal normalization distance measure $d_n(r)$.} (which is measured in the eigenvector coordinate system from the centre to the $n$th particle), where
\begin{equation}
	d_n^2(r) = x_{1,n}^2 + \frac{x_{2,n}^2}{q^2(r)} + \frac{x_{3,n}^2}{s^2(r)} \,.
\end{equation}

We then find the three eigenvalues $\left( \lambda_1 \ge \lambda_2 \ge \lambda_3 \right)$ of the matrix $\tilde{I}_{ij}$ and set the ellipsoidal orientation to the corresponding orthogonal eigenvectors $\left\{ \mathbf{e_1},\mathbf{e_2},\mathbf{e_3} \right\}$ (i.e. the principal axes).  The square roots of the eigenvalues are used to find the axis ratios: $s = \left( \lambda_3/\lambda_1 \right)^{1/2}$, $p = \left( \lambda_3/\lambda_2 \right)^{1/2}$, and $q = \left( \lambda_2/\lambda_1 \right)^{1/2}$.  The axis lengths are then computed with these axis ratios: $a(r)=r$, $b(r)=r\,q(r)$, and $c(r)=r\,s(r)$.  This ensures that the triaxial ellipsoid is contiguous to the bounding sphere of radius $r$ at two points.

For every $r$, the ellipsoid is initialized as a sphere, i.e. $s(r)=p(r)=q(r)=1$, and the inertia matrix, eigenvalues, and eigenvectors are computed.  In the second (and every subsequent) iteration, the inertia matrix is recomputed using particles that fall inside the reshaped and reoriented ellipsoid from the previous iteration.  This method keeps the largest axis length $a(r)$ constant, and thus, constrains this semi-major axis of the ellipsoid to lie on the surface of the bounding sphere.  Therefore, particles are added and removed to the set only within spherical radius $r$.  We continue the iterations until either $\Delta s=\left|s_k-s_{k-1}\right|$, $\Delta p=\left|p_k-p_{k-1}\right|$, and $\Delta q=\left|q_k-q_{k-1}\right|$ are all $< 10^{-3}$, or until a maximum of $k=1000$ iterations.

\section{Simulations of Milky-Way-Mass Galaxies}
\label{sec:sims}

This work compares different resimulations of three zoomed-in, cosmological-baryonic simulations of MW-mass haloes from the Latte suite of FIRE-2 simulations \citep[see][]{2018MNRAS.480..800H}.  The initial conditions for the haloes are all drawn from the same low-resolution cosmological box \citep[AGORA;][]{2014ApJS..210...14K} and are labelled \mf, \mi, and \mm.  The size of the zoomed-in region varies between $2$--$5$ Mpc, depending on the formation history of each halo.  The haloes are selected to have present-day virial masses between $1.2$--$1.6 \times 10^{12}$ M$_\odot$, similar to that of the MW, and to have no massive companions within at least $5 R_{200 \mathrm{m}} \sim 1.5$ Mpc.\footnote{$R_{200 \mathrm{m}}$ is the radius within which the total mass density,\\$\bar{\rho} \equiv 3M(<R_{200 \mathrm{m}})/4\pi R_{200 \mathrm{m}}^3$, is 200 times the average matter density.}

 The full FIRE-2 suite of 18 cosmological-baryonic zooms that have been run at this mass scale includes simulations that form thin discs similar to the MW's, as well as some that form spheroids \citep[][]{2018MNRAS.473.1930E, 2018MNRAS.481.4133G}.  We select these particular three systems for resimulation because the properties of their CDM versions have been extensively compared to the MW.  Their disc structure \citep[][]{2020ApJS..246....6S}, their satellite galaxy systems \citep[][]{2020arXiv201008571S}, and their stellar haloes \citep[][]{2018ApJ...869...12S} all have reasonable similarity to the MW.  These are not the only three systems within the suite for which this is true, but they are the ones for which the similarities and differences have been most thoroughly quantified.  As in the CDM versions, we use an initial mass resolution of 7100 M$_\odot$ for the star and gas particles, and 35000 M$_\odot$ for DM particles, for all resimulations.

The primary difference between the three haloes chosen for resimulation is their formation histories.  \mm\ forms earliest and has the largest disc scale radius of the three \citep[][]{2019MNRAS.485.5073D}.  \mf\ forms latest and includes a late interaction with a Small Magellanic Cloud (SMC)-mass galaxy that disrupts the disc \citep[][]{2018ApJ...869...12S}.  \mi\ has an intermediate formation epoch, the largest proportion of accretion from low-mass satellites, and a thicker, younger disc with a significant outer warp \citep[][]{2016ApJ...827L..23W, 2020ApJS..246....6S}.

We compare resimulations of these three haloes with the following set of variations at redshift $z=0$, summarized in Table \ref{tbl:resims}:

\begin{table*}
	\caption{\textbf{Summary of simulated MW-mass galaxy properties.}  All simulations have baryonic particle mass $\sim 7100$ M$_\odot$ and DM particle mass 35000 M$_\odot$.  An interaction cross-section of $\sigma/m=0$ cm$^2$ g$^{-1}$ indicates standard collisionless CDM.  $M_{\mathrm{vir}}, \ r_{\mathrm{vir}}$: \citet{1998ApJ...495...80B} spherical virial quantities.  $r_{-2}$: spherical radius where log-slope of DM density profile is $-2$.  $d_1$: scattering radius, determined as shown in Figure \ref{fig:figr_Gscatter} (\S\ref{subsec:dvr}); DNI indicates that $\Gamma_\mathrm{scatter}<t_{z=0}^{-1}$  for all $d$ with sufficient particles to determine $\rho_\mathrm{DM}(d)$ and $v_\mathrm{rel}(d) \approx 1.3 \, v_\mathrm{rms}(d)$ (see Figure \ref{fig:figr_Gscatter}); that is, the scattering rate profile ``does not intersect'' this characteristic value, so $d_1$ is undefined.  $M_{\star,90}, \ r_{\star,90}$: Mass and spherical radius of 90\% of stellar mass within 30 kpc of the central galaxy.  $\dag$: $d_1$ is determined using ellipsoidally-averaged DM local collision rate profiles from ellipsoidally-averaged density and velocity profiles; all other quantities in this table are determined with spherical averaging.  $\ddag$: Values for all simulations are taken at redshift $z=0$ except for \mi\ SIDM+Baryon $\sigma/m = 10$ cm$^2$ g$^{-1}$, which is evaluated at $z=0.1$.}
	\label{tbl:resims}
	\begin{tabular}{lcccccccl}
	
	    \hline
	    \hline
	    Initial Conditions &
		$\sigma/m$         &
		$M_{\mathrm{vir}}$ &
		$r_{\mathrm{vir}}$ &
		$r_{-2}$           &
		$d_1 \,^\dag$      &
		$M_{\star,90}$     &
		$r_{\star,90}$     &
		Reference
		\\[1pt]
		                                            &
		$\left[ \text{cm}^2 \text{ g}^{-1} \right]$ &
		$\left[ 10^{12} \ \text{M}_\odot \right]$   &
		$\left[ \text{kpc} \right]$                 &
		$\left[ \text{kpc} \right]$                 &
		$\left[ \text{kpc} \right]$                 &
		$\left[ 10^{10} \ \text{M}_\odot \right]$   &
		$\left[ \text{kpc} \right]$                 &
		\\
		\hline
	
    	\textbf{CDM-only}  \\
    	\mf           & 0   & 1.28 & 284.2 &   64.6 & --          & --  & --   & \citet{2017MNRAS.471.1709G} \\
    	\mi           & 0   & 0.90 & 252.8 &   24.5 & --          & --  & --   & \citet{2016ApJ...827L..23W} \\
    	\mm           & 0   & 1.14 & 273.9 &   35.5 & --          & --  & --   & \citet{2019MNRAS.487.1380G} \\
    	\textbf{SIDM-only} \\
    	\mf           & 1   & 1.28 & 284.0 &   38.9 & DNI         & --  & --   & \citet{2021arXiv210212480S} \\
    	\mf           & 10  & 1.25 & 282.0 &   28.2 &        24.  & --  & --   & \citet{2021arXiv210212480S} \\
    	\textbf{CDM+Baryon}  \\
    	\mf           & 0   & 1.33 & 287.9 &   19.5 & --          & 5.3 & 12.4 & \citet{2017MNRAS.471.1709G} \\
    	\mi           & 0   & 0.96 & 258.4 &   17.0 & --          & 3.2 & 16.7 & \citet{2016ApJ...827L..23W} \\
    	\mm           & 0   & 1.23 & 280.8 &   18.6 & --          & 4.9 & 21.2 & \citet{2018MNRAS.480..800H} \\
    	\textbf{SIDM+Baryon} \\
    	\mm           & 0.1 & 1.22 & 279.9 &   17.0 & DNI         & 6.1 & 23.0 & \citet{2021arXiv210212480S} \\
    	\mf           & 1   & 1.36 & 289.8 &   13.5 & \,\;\;  8.8 & 6.2 & 15.7 & \citet{2021arXiv210212480S} \\
    	\mi           & 1   & 0.98 & 260.0 &   10.7 & \,\;\;  7.4 & 5.0 & 13.9 & \citet{2021arXiv210212480S} \\
    	\mm           & 1   & 1.24 & 281.5 & \, 9.8 & \,\;\;  9.8 & 6.6 & 20.2 & \citet{2021arXiv210212480S} \\
    	\mf           & 10  & 1.27 & 283.2 &   28.2 &        23.  & 5.1 & 15.9 & \citet{2021arXiv210212480S} \\
    	\mi$\,^\ddag$ & 10  & 0.92 & 237.8 &   24.5 &        19.  & 4.5 & 12.3 & \citet{2021arXiv210212480S} \\
    	\mm           & 10  & 1.22 & 279.6 & \, 5.4 &        22.  & 8.0 & 20.4 & \citet{2021arXiv210212480S} \\
    	\hline
    	\hline
    	
	\end{tabular}
\end{table*}

\begin{enumerate}
    \item \textbf{CDM-only -- Collisionless CDM without baryons}, for all three haloes;
    \item \textbf{SIDM-only -- Collisional self-interacting DM without baryons}, at $ \sigma / m = $ 1 and 10 cm$^2$ g$^{-1}$ for \mf;
    \item \textbf{CDM+Baryon -- Collisionless CDM with baryons and full hydrodynamics}, using FIRE-2 feedback recipes, for all three haloes;
    \item \textbf{SIDM+Baryon -- Collisional self-interacting DM with baryons and full hydrodynamics}, with identical baryonic physics to the fiducial suite, at $\sigma/m=0.1$ cm$^2$ g$^{-1}$ for \mm, and 1 and 10 cm$^2$ g$^{-1}$ for all three haloes, but at redshift $z=0.1 \ \left( t=12.5 \ \textrm{Gyr} \right)$ for \mi\ at the latter cross-section.
\end{enumerate}
All DM self-interactions are realized using Monte Carlo elastic (non-dissipative) scattering, as described in \citet{2013MNRAS.430...81R}.  We evaluate the \mi\ SIDM+Baryon $\sigma/m =$ 10 cm$^2$ g$^{-1}$ at $z=0.1$, the latest epoch currently available for this resimulation.  Based on the behavior observed in the other runs, we expect the radial density and velocity profile of this simulated galaxy to be relatively stable between $z=0.1$ and $z=0$.  The shape profiles of the various species continue to evolve to $z=0$ in the central parts of the galaxies, but this effect is least pronounced for the DM component.

The baryonic runs listed above and in Table \ref{tbl:resims} use the standard set of FIRE-2 feedback recipes \citep[][]{2018MNRAS.480..800H} with one exception, which is to ignore the conversion of thermal to kinetic energy in the unresolved Sedov-Taylor phase of the expansion of shocks generated by mass loss from massive stars.  As discussed in \citep[][]{2021arXiv210212480S}, the default prescription in FIRE-2 had the effect of converting nearly all the thermal energy into momentum, giving the stellar winds a similar effect to a miniature supernova and resulting (perhaps counter-intuitively) in higher star formation rate (SFR) and stellar mass in the simulated galaxies, and subsequently less diversity among density profiles.  However, for this study we use the resimulations of the CDM haloes that ignore this ``$PdV$'' work for the sub-res regions, for consistency with the existing SIDM runs. 

\begin{figure*}
    \centering
    \includegraphics[width =  \textwidth,
                     height = \textheight,
	                 keepaspectratio]
    {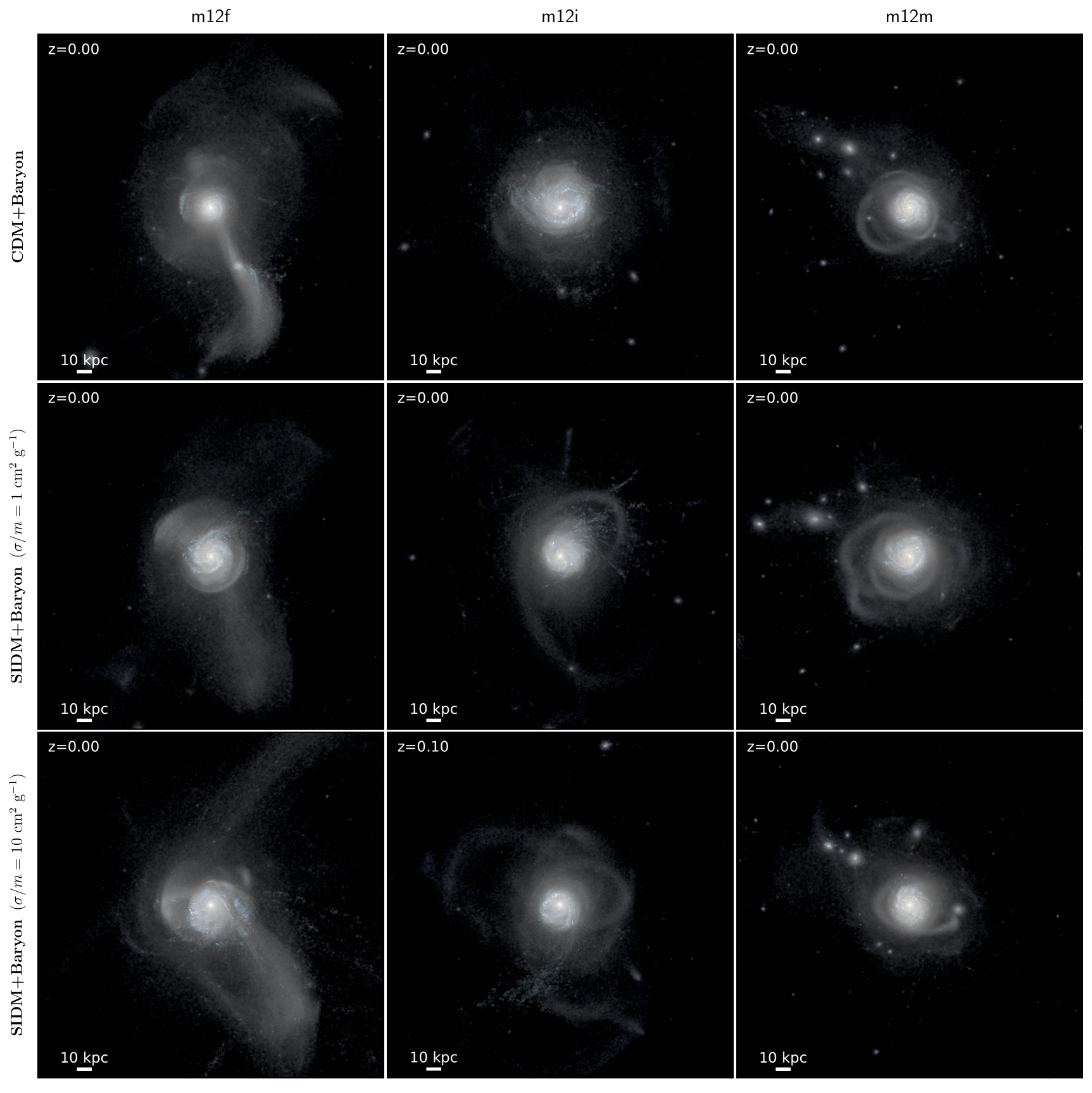}
    \caption{\textbf{Face-on halo view of cosmological-baryonic simulated MW-mass galaxies.}  SED-modeled, ray-traced images of starlight in three sets of MW-mass galaxy simulations: \mf\ (left), \mi\ (centre), and \mm\ (right); all DM+Baryon simulations are shown except \mm\ SIDM+Baryon $\sigma/m=0.1$ cm$^2$ g$^{-1}$ (which is quite similar to the \mm\ CDM+Baryon case).  Each panel is 300 kpc across and the galaxy has been rotated to show the disc face-on.  All simulations except \mi\ SIDM+Baryon $\sigma/m=10$ cm$^2$ g$^{-1}$ (discussed in \S\ref{sec:sims}) are depicted at $z=0$.}
    \label{fig:figr_faceon_halo}
\end{figure*}

Figure \ref{fig:figr_faceon_halo} shows Spectral Energy Distribution (SED)-modeled and ray-traced images (integrated over filter band-passes similar to Hubble Space Telescope visible light) of the simulated galaxies in a wide-angle view focused on the stellar halo.  It is apparent that although there are some small differences, every simulation contains a reasonable-looking spiral galaxy, system of dwarf satellite galaxies, and stellar halo.  One obvious difference between resimulations of the same initial conditions is in the particular dwarf satellites that appear in each simulation, which vary between even identical runs due to the stochastic impact of supernovae.  The most massive satellites are stable to this effect but can have small phase differences in their orbits from run to run, as is evident in \mf.  The other obvious difference is in the star formation rate (as evident in \mi).  Star formation is a highly non-linear process, and also varies stochastically from run to run even for identical initial conditions, again mainly because of the random occurrence and clustering of supernovae and also (as clearly seen in \mf) from differences in the orbital phase of mergers.  However, these differences do not significantly change the global properties of the haloes and their central galaxies, as we next demonstrate.

\begin{figure*}
    \centering
    \includegraphics[width  = \textwidth,
                     height = \textheight,
	                 keepaspectratio]
    {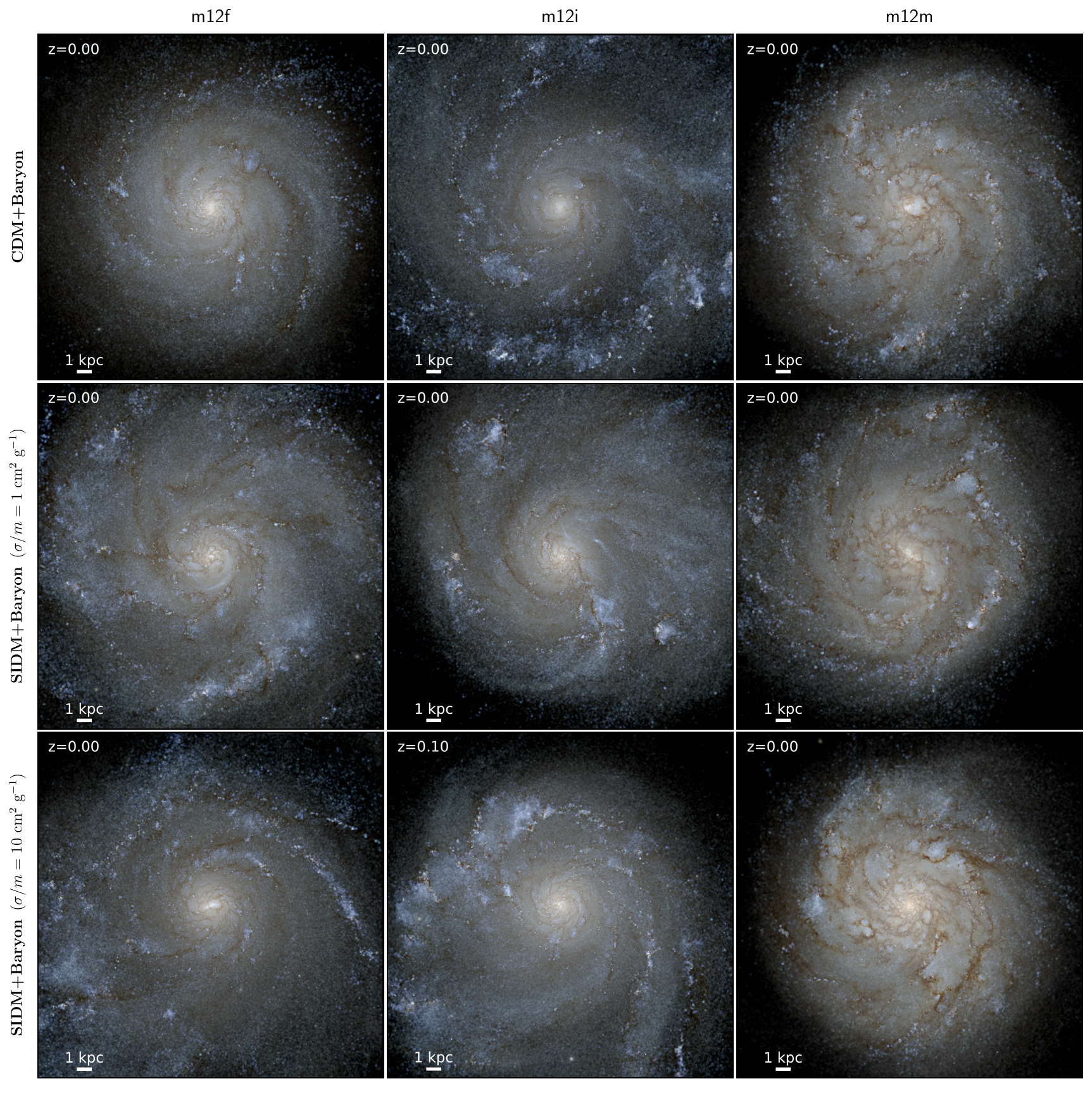}
    \caption{\textbf{Face-on disc view of cosmological-baryonic simulated MW-mass galaxies.}  SED-modeled, ray-traced images of starlight in three sets of MW-mass galaxy simulations: \mf\ (left), \mi\ (centre), and \mm\ (right); all DM+Baryon simulations are shown except \mm\ SIDM+Baryon $\sigma/m=0.1$ cm$^2$ g$^{-1}$ (which is quite similar to the \mm\ CDM+Baryon case).  Each panel is 30 kpc across and the galaxy has been rotated to show the disc face-on.  All simulations except \mi\ SIDM+Baryon $\sigma/m=10$ cm$^2$ g$^{-1}$ (discussed in \S\ref{sec:sims}) are depicted at $z=0$.}
    \label{fig:figr_faceon_disk}
\end{figure*}

Figure \ref{fig:figr_faceon_disk} also shows SED-modeled and ray-traced images, but with close-up views of the stellar discs in the simulated systems.  There is remarkable uniformity in the structure and size of the discs across all DM simulations.  The \mf\ SIDM simulation has a slightly higher SFR in its outskirts, probably due to the timing of a merger with a roughly $\sim$SMC-mass object visible in Figure \ref{fig:figr_faceon_halo}, while \mi\ and \mm\ have no noticeable increase in star formation with $\sigma/m$.  Generally, the discs of the CDM galaxies tend to be slightly more massive and compact than in SIDM.

\begin{figure*}
	\centering
	\includegraphics[width  = \textwidth,
	                 height = \textheight,
	                 keepaspectratio]
	{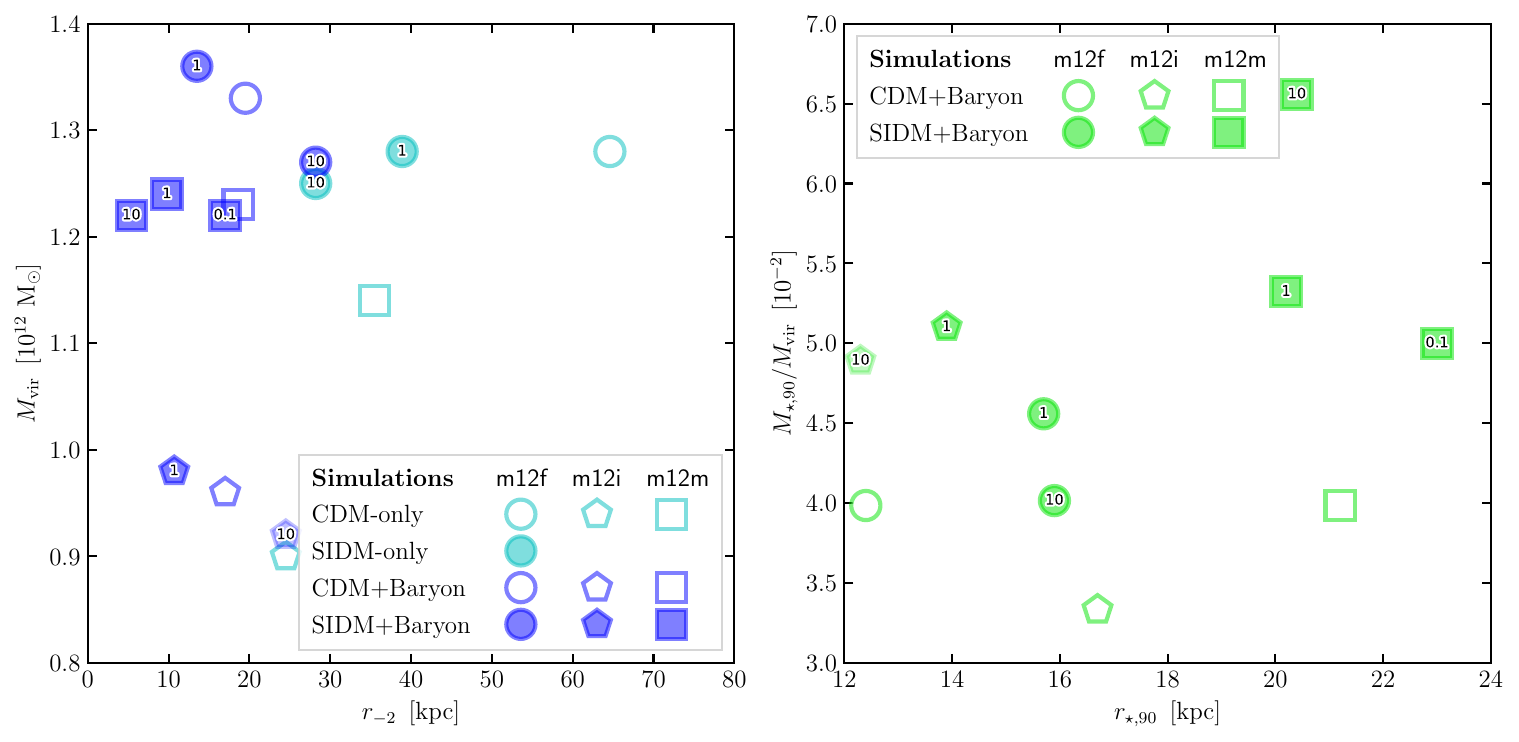}
	\caption{\textbf{Global DM and stellar properties of simulated MW-mass galaxies.} DM virial mass $M_\mathrm{vir}$ as a function of DM scale radius $r_{-2}$ (left) and stellar-to-halo-mass ratio $M_{\star,90}/M_\mathrm{vir}$ as a function of radius enclosing 90\% of stellar mass (right) for \mf\ (circles), \mi\ (pentagons), and \mm\ (squares).  In the left-hand panel (DM properties), the CDM-only, SIDM-only, CDM+Baryon, and SIDM+Baryon simulations are represented with hollow cyan, solid cyan, hollow blue, and solid blue markers, respectively; in the right-hand panel (stellar properties), the CDM+Baryon and SIDM+Baryon simulations are represented with hollow green and solid green markers, respectively. In both panels, the SIDM+Baryon simulations with $\sigma/m = 0.1$, 1, and 10 cm$^2$ g$^{-1}$ are shown with these respective numbers inside the markers.  The simulation of \mi\ SIDM+Baryon at $\sigma/m = 10$ cm$^2$ g$^{-1}$ is evaluated at $z=0.1$ (instead of $z=0$ like all other simulations), indicated with a decrease in alpha (lighter shade).}
	\label{fig:figr_Mvr}
\end{figure*}

Figure \ref{fig:figr_Mvr} summarizes the characteristic masses and radii of the resimulations, quantifying the impressions given by examining Figures \ref{fig:figr_faceon_halo} and \ref{fig:figr_faceon_disk}.  The left-hand panel shows the DM virial mass $M_\mathrm{vir}$ versus DM scale radius $r_{-2}$; the right-hand panel shows the  stellar-to-halo-mass ratio $M_{\star,90}/M_\mathrm{vir}$ versus the radius enclosing 90\% of the stellar mass $r_{\star,90}$.  These values are all computed using spherical volumes.  $M_\mathrm{vir}$ values are roughly the same over scattered domains of $r_{-2}$ for each set of simulations, while the $M_{\star,90}/M_\mathrm{vir}$ ratios are generally more scattered over $r_{\star,90}$.  The CDM-only and SIDM-only simulations have larger $r_{-2}$ than their CDM+Baryon and SIDM+Baryon counterparts due to the increased concentration produced by the central baryonic component.  For CDM+Baryon and SIDM+Baryon, the \mf\ and \mi\ simulations have the smallest and largest $M_\mathrm{vir}$, respectively, while the \mm\ simulations fall in-between.  The interaction cross-section does not otherwise seem to produce any clear trends in the global DM distribution; \mm's scale radius decreases as $\sigma/m$ increases, while \mf\ and \mi\ show no clear trend.  In all cases the virial mass varies by less than 10\% across all cross-sections.  Finally, steady growth in $M_{\rm vir}$ and $M_{\star,90}$ for \mi\ at 10 cm$^2$ g$^{-1}$ from $0.1$ (plotted) to $z=0$ (all other runs) would bring this halo into consistency with the other \mi\ simulations, as would further contraction of the DM distribution due to the baryonic component (leading to a decrease in $r_{-2}$).

In terms of the stellar distributions, there is significant variation in the stellar-to-halo mass ratio across the different resimulations, while $r_{\star,90}$ appears roughly independent of $\sigma/m$ for most cases (although $r_{\star,50}$ does have a trend with cross section; see \citealt{2021arXiv210212480S}). Again \mm\ shows the opposite trend from \mi\ and \mf, as well as the largest variation in $M_{\star,90}/M_\mathrm{vir}$.  While in \mf\ and \mi\, $M_{\star,90}/M_\mathrm{vir}$ shows no trend for larger $\sigma/m$, for \mm\, a larger $\sigma/m$ gives rise to a relatively more massive central galaxy (recall that the DM halo does not change appreciably in mass between runs).

\section{Results}
\label{sec:results}

\begin{figure}
	\centering
	\includegraphics[width  = \columnwidth,
	                 height = \textheight,
	                 keepaspectratio]
	{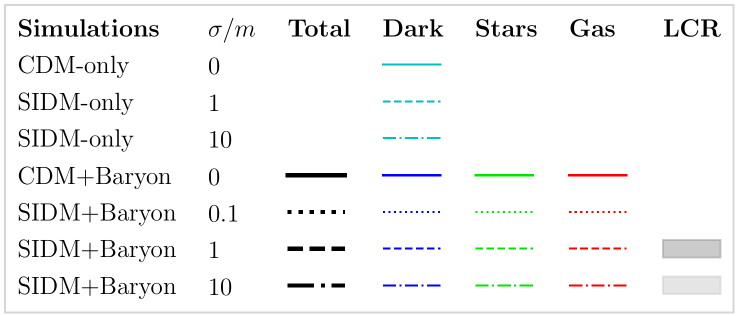}
	\caption{\textbf{Legend for comparisons of simulated MW-mass galaxies in CDM and SIDM.}  Line-styles denote different DM cross-sections listed above in units of cm$^2$ g$^{-1}$, colours show the different species, and shaded-areas indicate the SIDM local collision regions (LCR).  These line-styles, colours, and shaded-areas are used throughout the rest of this work and the Appendix.}
	\label{fig:figr_legend}
\end{figure}

In this \S, we compare the density, velocity, and shape profiles of the different simulations described in \S \ref{sec:sims}.  Throughout the rest of this work, we use the same series of line-styles to denote different DM cross-sections, different colours to show the different species, and gradient shaded-areas to indicate the SIDM local collision regions (LCR) for different $\sigma/m$; these are given in Figure \ref{fig:figr_legend} and apply to Figures \ref{fig:figr_dens} through \ref{fig:figr_Delta_s_vs_r}, as well as Figures \ref{fig:figr_GscatterES}, \ref{fig:figr_ca_all_geometricR}, and \ref{fig:figr_T_geometricR} in the Appendix.

\subsection{Densities, Velocities, and Scattering Rates}
\label{subsec:dvr}

Predictions for density and velocity distributions in SIDM are closely related to those for the halo's shape, since the same interactions that heat the inner regions also make the outer halo more spherical, by preferentially scattering DM particles on plunging, radial orbits.  Before discussing the shapes, we will review the density and velocity profiles for DM \citep[discussed in full in][]{2021arXiv210212480S} and discuss the profiles for the stellar and gas components as well.  While \citet{2021arXiv210212480S} presents spherically-averaged profiles, here we show profiles computed using ellipsoids fit to isodensity contours for each species (\S \ref{sec:methods}).  The difference in density, and therefore DM scattering rate, can vary by up to 60\% from the spherically-averaged value, depending on the flattening (Appendix \ref{sec:evs}).

To compute the density $\rho$, we use the mass enclosed in shells that follow the triaxial ellipsoidal surfaces calculated using the method described in \S\ref{sec:methods}, which approximately follow isodensity surfaces.  However, we use much broader spacing in distance for the density calculation than for the ellipsoid fits, spacing shells by roughly every 10th point in $r$ for which a fit is carried out, to allow enough space between shells to get sufficient numbers of particles and to mitigate problems caused by the twisting of the ellipsoid axes between shells.  To estimate the density at ellipsoidal distance $d_\kappa$, we select all particles $N_\kappa$ within a shell $\kappa$ centered on the ellipsoid with semi-major axis $a_{\kappa}$ (discarding shells with $N_\kappa<100$).  The shell half-thickness $\Delta d_\kappa$ is set by the difference in semi-major axis between the isodensity surface at $d_\kappa$ and its inner neighbor, such that $\Delta d_\kappa \equiv a_\kappa - a_{\kappa-1}$.  Then the density $\rho(d_\kappa)$ is computed as
\begin{equation}
    \rho(d_\kappa) \equiv \frac{\sum_{n=1}^{N_\kappa} m_n }{ \frac{4}{3} \pi q_\kappa s_\kappa \left[ \left(d_\kappa +\Delta d_\kappa\right)^3 -  \left(d_\kappa -\Delta d_\kappa\right)^3 \right] } \,,
\end{equation}
where $q_\kappa$ and $s_\kappa$ are the axis ratios of the ellipsoid used to calculate $d_{\kappa}$.  Likewise the root-mean-square (RMS) velocity $v_\mathrm{rms}(d_\kappa)$ is
\begin{equation}
    v_\mathrm{rms}(d_\kappa) \equiv \sqrt{\frac{1}{N_\kappa} \sum_{n=1}^{N_\kappa} \mathbf{v}_n \cdot \mathbf{v}_n } \,,
\end{equation}
where $\mathbf{v}_n$ is the velocity vector of particle $n$ inside shell $\kappa$.

\begin{figure*}
	\centering
	\includegraphics[width  = \textwidth,
	                 height = \textheight,
	                 keepaspectratio]
	{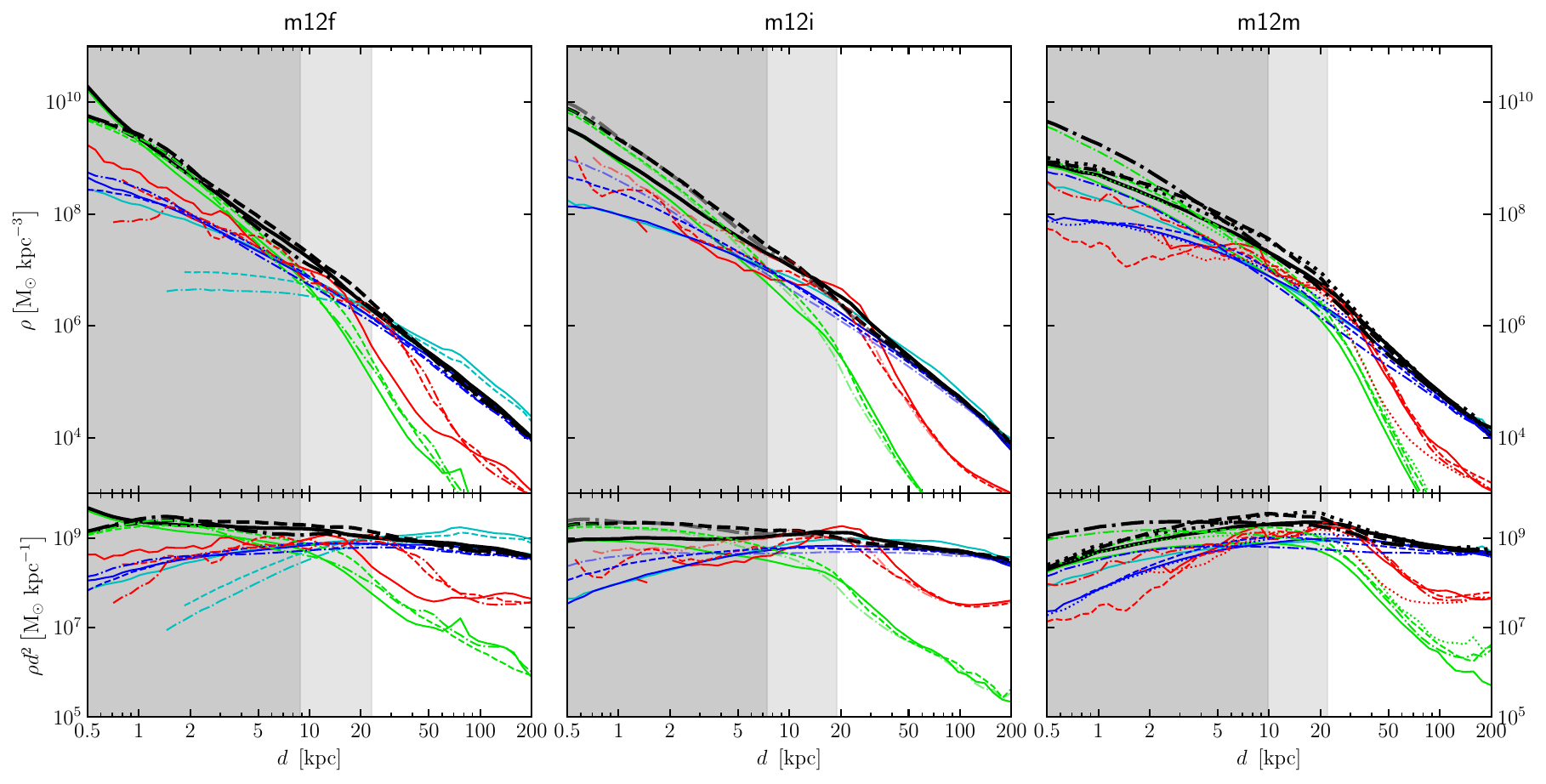}
	\caption{\textbf{Ellipsoidally-averaged density profiles of the simulated MW-mass galaxies.}  Density $\rho$ (top) and $\rho d^2$ (bottom; an ideal isothermal profile is flat in this view) as a function of ellipsoidal distance $d$ for \mf\ (left), \mi\ (centre), and \mm\ (right).  Line-styles, colours, and shaded-areas follow the legend in Figure \ref{fig:figr_legend}.}
	\label{fig:figr_dens}
\end{figure*}

\begin{figure*}
	\centering
	\includegraphics[width  = \textwidth,
	                 height = \textheight,
	                 keepaspectratio]
	{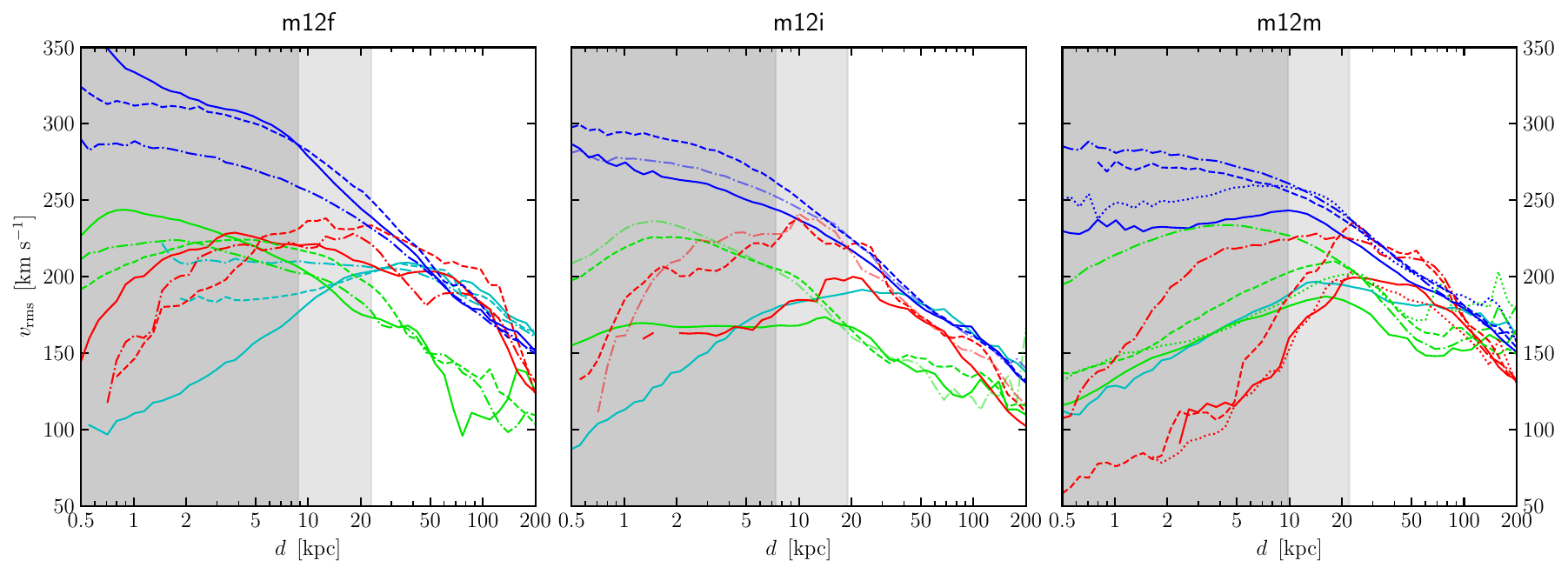}
	\caption{\textbf{Ellipsoidally-averaged velocity profiles of the simulated MW-mass galaxies.}  Root-mean-square velocity $v_\mathrm{rms}$ as a function of ellipsoidal distance $d$ for \mf\ (left), \mi\ (centre), and \mm\ (right).  Line-styles, colours, and shaded-areas follow the legend in Figure \ref{fig:figr_legend}.}
	\label{fig:figr_vrms}
\end{figure*}

We also calculate the DM local collision rate:
\begin{align}
	\Gamma_\mathrm{scatter}(d) & =  \rho_\mathrm{DM}(d) \, v_\mathrm{rel}(d) \, \sigma/m \nonumber\\
	                           &\approx \frac{4}{\sqrt{3\pi}} \, \rho_\mathrm{DM}(d) \, v_\mathrm{rms}(d) \, \sigma/m \,, \label{eqn:R_scatter}
\end{align}
where $\rho_\mathrm{DM}(d)$ is the DM density and $v_\mathrm{rel}(d)$ is the relative DM velocity.  We approximate $v_\mathrm{rel}$ by $4/\sqrt{3\pi} \, v_\mathrm{rms} \approx 1.3 \, v_\mathrm{rms}$, which holds exactly for a Maxwellian velocity distribution. We define the local collision region (LCR) scattering radius $d_1$ by requiring $\Gamma_\mathrm{scatter}(d_1) =  t_{z=0}^{-1}$, where $t_z$ is the time at redshift $z$, thus $t_{z=0}=13.8$ Gyr is age of the Universe \citep[][Table A.1: \textit{Planck}+WP+BAO]{2014A&A...571A..16P}.\footnote{\mi\ SIDM+Baryon $\sigma/m = 10$ cm$^2$ g$^{-1}$ is evaluated at $z=0.1$, which gives $\Gamma_\mathrm{scatter}(d_1)=t_{z=0.1}^{-1}$ where $t_{z=0.1}=12.5$ Gyr.}  Thus, inside $d_1$, DM particles have experienced at least one self-scatter within the age of the Universe (based on the density distribution at $z=0$), giving an approximate volume inside which we expect the SIDM differences to be the largest.  The $d_1$ for different simulations are given in Table \ref{tbl:resims}, and the LCR for different SIDM+Baryon $\sigma/m$ are represented as gray shaded-areas (as shown by the legend in Figure \ref{fig:figr_legend}) and used in Figures \ref{fig:figr_dens}, \ref{fig:figr_vrms},  \ref{fig:figr_ca_all_sphericalr}, \ref{fig:figr_T_sphericalr}, \ref{fig:figr_Delta_s_vs_r}, \ref{fig:figr_ca_all_geometricR}, and \ref{fig:figr_T_geometricR}.

As expected, the SIDM-only density profiles for \mf\ (Figure \ref{fig:figr_dens}, left) have much lower densities in the central region compared to all the other DM profiles, while the DM profiles in all the CDM-only, CDM+Baryon, and SIDM+Baryon simulations are remarkably similar.  The CDM-only and SIDM-only densities are too low to produce smooth curves for $r\lesssim 2$ kpc.

In the simulations with baryons, the density of stars dominates the central region, while the DM and gas approximately follow each other about a magnitude below the stellar component.  The stellar mass density of \mm\ is lower in the central region and higher at larger radii compared to that of \mf\ and \mi.  Notably, when examining the bottom row showing $\rho d^2$, there does not appear to be a transition at $d_1$ from an isothermal (flat in this view) profile at $r<d_1$ to a Navarro-Frenk-White (NFW)-like profile at $r>d_1$, as posited in \citet{2000PhRvL..84.3760S}.  This is another illustration of the effect of the growing galaxy in the centre in altering the density profile well beyond the baryon-dominated region, thanks to the significant radial anisotropy in the DM velocity distribution.

Root-mean-square (RMS) velocity profiles are shown in Figure \ref{fig:figr_vrms}.  The CDM-only and SIDM-only simulations have much lower central RMS velocities than the corresponding DM curves in the CDM+Baryon and SIDM+Baryon simulations, except in the extreme outskirts ($r \gtrsim 100$ kpc).  The additional component added by the baryons deepens the central potential and subsequently increases the DM density, leading to higher RMS velocities.  The biggest differences are between CDM-only and CDM+Baryon in the central region for \mf\ and \mi.  The velocity dispersion in the CDM+Baryon case is much higher than for CDM-only in both these simulations due to baryonic deepening of the potential.  In the SIDM case, the effects of baryonic contraction are offset by the ability of the DM to thermalize; that is, to transfer some of the energy outward that would otherwise go into raising the central velocity dispersion.  Surprisingly, for \mm, the SIDM+Baryon runs have significantly larger central velocity dispersion than the CDM+Baryon run does.  This is probably related to the fact that the central galaxy in \mm\ is substantially more massive as $\sigma/m$ increases without becoming significantly larger in extent (Figure \ref{fig:figr_Mvr}), implying that the average stellar density is larger in the SIDM+Baryon runs than the CDM+Baryon one.  Indeed, the central galaxy in \mi\ shows the same tendency for the SIDM systems to be slightly more massive in stellar mass, but has slightly less variation between DM models than \mm.  This variety illustrates how the central density and velocity dispersion of the DM are shaped by interplay with the growing galaxy in the center of the halo.

\begin{figure}
	\centering
	\includegraphics[width  = \columnwidth,
	                 height = \textheight,
	                 keepaspectratio]
	{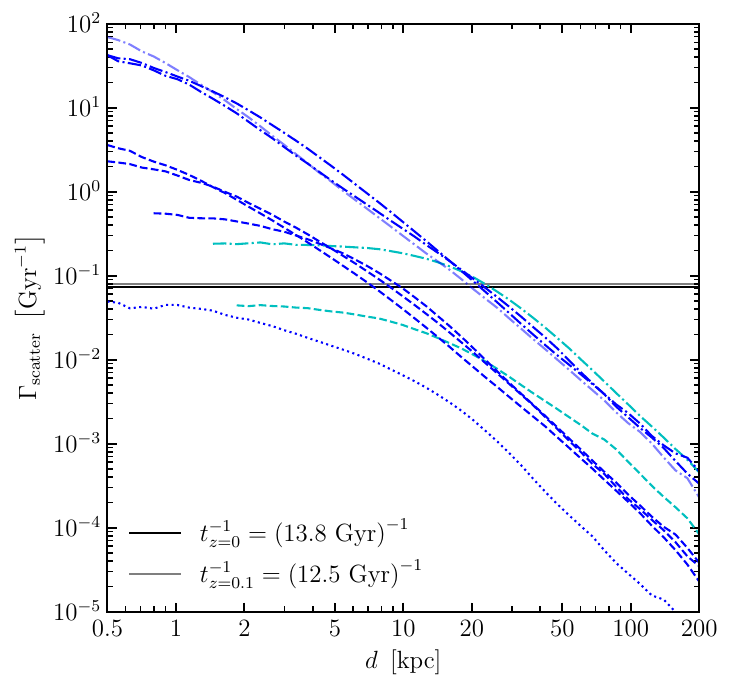}
	\caption{\textbf{Ellipsoidally-averaged DM local collision rate profiles of the simulated MW-mass SIDM galaxies.}  The local DM self-scattering rate $\Gamma_\mathrm{scatter}$ (Equation \ref{eqn:R_scatter}) is shown as a function of ellipsoidal distance $d$ for SIDM simulations.  Line-styles and colours follow the legend in Figure \ref{fig:figr_legend}.  The point where $\Gamma_{\mathrm{scatter}}$ intersects the gray horizontal lines at $t_{z=0}^{-1}$ and $t_{z=0.1}^{-1}$ marks the scattering radius $d_1$, shown as the shaded-areas of Figures \ref{fig:figr_dens}, \ref{fig:figr_vrms}, \ref{fig:figr_ca_all_sphericalr}, \ref{fig:figr_T_sphericalr}, \ref{fig:figr_Delta_s_vs_r}, \ref{fig:figr_ca_all_geometricR}, and \ref{fig:figr_T_geometricR}.
	}
	\label{fig:figr_Gscatter}
\end{figure}

Figure \ref{fig:figr_Gscatter} shows the local collision rate $\Gamma_\mathrm{scatter}$ as a function of ellipsoidal distance $d$.  The scattering radius $d_1$ is smaller for $\sigma/m = 1$ cm$^2$ g$^{-1}$ than 10 cm$^2$ g$^{-1}$, while the rate for 0.1 cm$^2$ g$^{-1}$ is less than $t_{z=0}^{-1}$ at all radii for the $z=0$ density distribution (though this is not necessarily true at all $z$).  The profile for \mf\ SIDM-only $\sigma/m = 1$ cm$^2$ g$^{-1}$ also does not reach $t_{z=0}^{-1}$, while 10 cm$^2$ g$^{-1}$ does.  This indicates that while self-interactions may have occurred earlier, the resulting heating reduces the scattering rate by $z=0$ to less than $t_{z=0}^{-1}$ everywhere.  On the other hand, all SIDM haloes with baryons have substantially higher interaction rates at the present day, likely as a result of the additional depth in the gravitational potential created by the central galaxy.  The growing galaxy can thus amplify the effect of a nonzero SIDM cross-section in the central portion of the halo by keeping the scattering rate higher over time.

\subsection{Shape Profiles}

\begin{figure*}
	\centering
	\includegraphics[width  = \textwidth,
	                 height = \textheight,
	                 keepaspectratio]
	{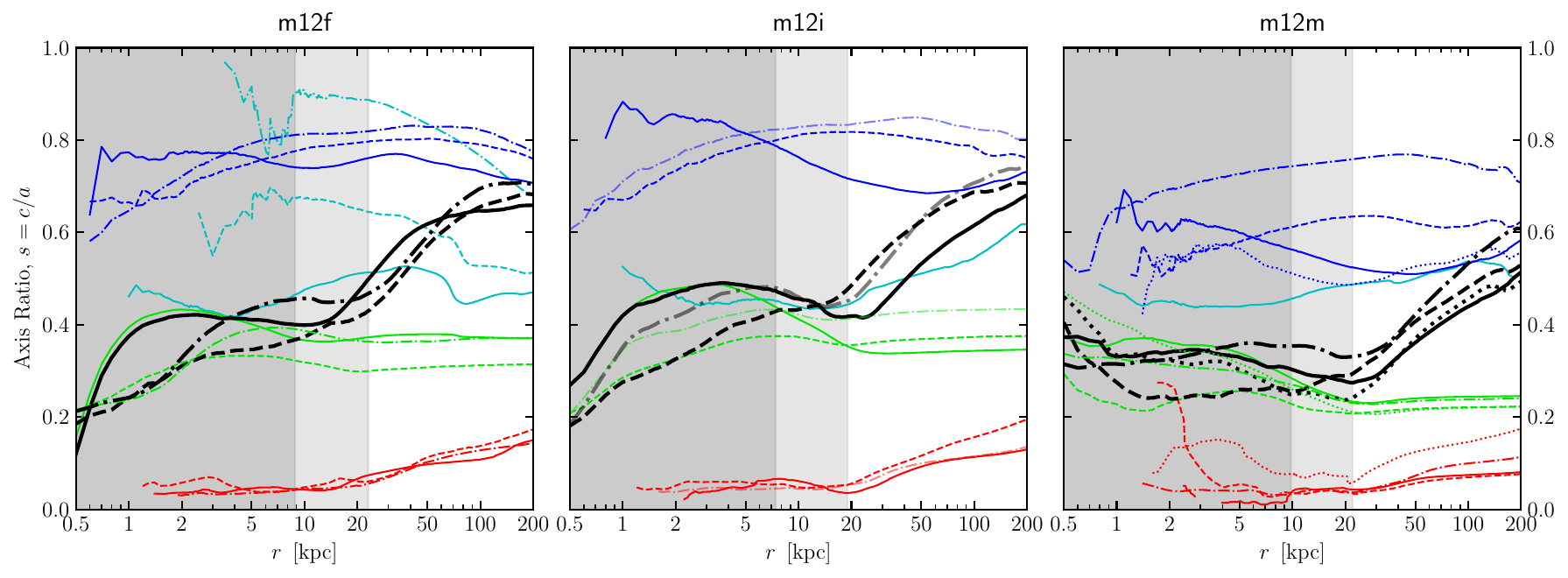}
	\caption{\textbf{Shape profiles of simulated MW-mass galaxies.} Minor-to-major axis ratio $s=c/a$ is shown as a function of semi-major axis $r=a$ for different species (DM, stars, and gas) in three sets of MW-mass galaxy simulations: \mf\ (left), \mi\ (centre), and \mm\ (right).  Line-styles, colours, and shaded-areas follow the legend in Figure \ref{fig:figr_legend}.  An alternate version of this plot using the geometrical mean of the axis lengths, $R = \left(abc\right)^{1/3}$, can be found in Appendix \ref{sec:geo}.}
	\label{fig:figr_ca_all_sphericalr}
\end{figure*}

The profiles for the minor-to-major axis ratio, $s=c/a$, for the different haloes in the simulations are shown in Figure \ref{fig:figr_ca_all_sphericalr}.  As summarized in Figure \ref{fig:figr_legend}, different colours distinguish between simulations with and without baryons and among species (DM, stars, and gas), while line-styles show different DM interaction cross-sections $\sigma/m$.  Axis ratios are calculated using the method described in \S \ref{sec:methods}, and we remove all data points where the triaxial ellipsoids do not enclose at least 5000 particles \citep[see Appendix A of][where  at least 3000 particles are used]{2011MNRAS.416.1377V}.

Axis ratio differences between the CDM+Baryon and SIDM+Baryon simulations are small compared to the differences between the CDM-only and SIDM-only simulations over the same range of cross-sections (Figure \ref{fig:figr_ca_all_sphericalr}).  SIDM-only creates the most spherical DM haloes, obtained for the highest $\sigma/m$.  Adding baryons to these simulations increases the overall roundness, but increasing $\sigma/m$ does not produce the large changes seen in the non-baryon cases.  The effect of the growth of the central galaxy is to standardize the shape in roughly the same range of axis ratios, between approximately 0.6 and 0.8.  In \mf\ (left) and \mi\ (centre), most of the scaling of the shape with cross-section is also erased.  In \mm\ (right) there are still substantial differences between haloes with different cross-sections, but the variation in shape has been ``recentered'' around the CDM case, while with DM-only the shape just gets progressively rounder for larger cross-sections.

In the central region of these MW-mass galaxies ($r \lesssim 10$ kpc), CDM+Baryon produces a more spherical DM distribution than SIDM+Baryon in nearly all cases.  This is consistent with, though less pronounced than, the effects discussed in \citet{2018MNRAS.479..359S}, but contrary to expectations from analytic predictions considering only DM \citep[][]{2018PhR...730....1T}.  The \emph{stellar} distribution is also frequently flatter for SIDM than CDM across all three galaxies.  These differences are greatest in the region where SIDM is collisional (gray shaded-areas), and the degree of flattening in the DM distribution parallels the flattening in the stars, indicating ongoing dynamical coupling between the stellar and DM distributions in the inner galaxy.

It is also apparent from the shape curves, which stop when the ellipsoid no longer encloses at least 5000 particles, that the SIDM haloes with baryons remain much denser in their centers at late times than those without baryons (as discussed in \S\ref{subsec:dvr}).  The LCR gray shaded-areas are calculated from the \emph{present-day} DM densities (Figure \ref{eqn:R_scatter}), indicating that while the SIDM-only haloes have reached an equilibrium where even their innermost regions have a very low collision rate relative to the age of the Universe, the gradually increasing potential depth due to the central galaxy counteracts collisional heating and maintains a much higher central collision rate at late times.  This also supports the idea that there is ongoing information exchange between the DM and stars in the inner galaxy, and suggests that this region may not be in equilibrium between the two species.

The \mm\ series of simulations contains most of the exceptions to these generalizations and is thus worth discussing in more detail.  This galaxy is the earliest of the three to form and has the largest scale radius (Figure \ref{fig:figr_Mvr}), meaning that its baryonic component has had the longest time to shape the DM distribution (and vice-versa) over the largest range of radii.  As pointed out in \S\ref{sec:sims}, its global properties have the opposite trend with $\sigma/m$ from the other two haloes.  It is also the only halo whose outer shape ($\gtrsim 20$ kpc) is consistent between the CDM-only and CDM+Baryon cases, and the only one where there is significant variation of the shape with cross-section across all radii.  Its particular assembly history (early accretion of many small galaxies) thus appears particularly sensitive to SIDM effects.  This could be because it simply has more time to establish equilibrium between the SIDM and stellar components, and is driven there more rapidly by a central relatively large galaxy that forms early.

A full comparison of the central DM shapes in these simulations is challenging, since many of the profiles are noisy due to low particle number.  This is especially true for the \mf\ SIDM-only simulations, whose shape profiles inside $5$--$10$ kpc are ambiguous thanks to their low central densities.  In all the DM+Baryon simulations, the shape of the stellar component is well-resolved (at all radii) and closely follows the shape of the \emph{total} mass distribution (at lower radii), which is accessible through dynamical modeling.  We will examine in more detail whether the differences in the three-dimensional (3D) shape of the \emph{stellar} distribution are observable in future work.

\subsection{Triaxiality}

\begin{figure*}
	\centering
	\includegraphics[width  = \textwidth,
	                 height = \textheight,
	                 keepaspectratio]
	{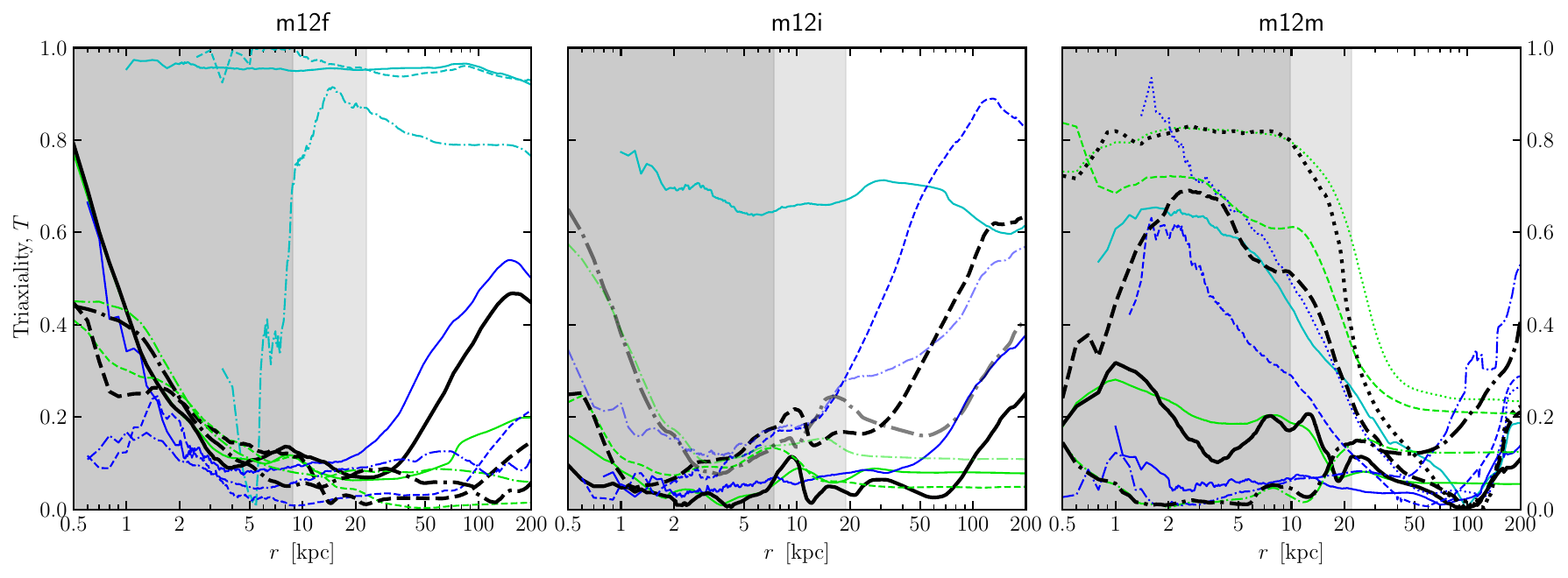}
	\caption{\textbf{Triaxiality profiles of simulated MW-mass galaxies.}  Triaxiality $T$ is shown as a function of semi-major axis $r=a$ for different species (DM and stars) in three sets of MW-mass galaxy simulations: \mf\ (left), \mi\ (centre), and \mm\ (right).  Line-styles, colours, and shaded-areas follow the legend in Figure \ref{fig:figr_legend}.  An alternate version of this plot using the geometrical mean of the axis lengths, $R = \left(abc\right)^{1/3}$, can be found in Appendix \ref{sec:geo}.}
	\label{fig:figr_T_sphericalr}
\end{figure*}

To better understand the shapes, we also calculate the triaxiality parameter $T$ \citep[][]{1991ApJ...383..112F}:
\begin{equation}
    T \equiv  \frac{a^2-b^2}{a^2-c^2} = \frac{1-q^2}{1-s^2} \,,
\end{equation}
where an ellipsoid is oblate if $0<T<1/3$ ($T=0$ is a perfect oblate distribution, $c \ll a=b$), triaxial if $1/3<T<2/3$ ($T=0.5$ is a maximally triaxial distribution), and prolate if $2/3<T<1$ ($T=1$ is a perfect prolate distribution, $b=c \ll a$).  Figure \ref{fig:figr_T_sphericalr} shows the triaxiality for the stellar and DM components of all simulations, as well as the total mass distribution.  There are a wide variety of behaviors on display.  Consistent with our other results, \mm\ has significantly different behavior than \mf\ and \mi.

The DM-only haloes for all three set of simulations are highly prolate or triaxial in the centre, especially \mf.  For CDM, this triaxiality/prolateness is well understood.  Interestingly, though, at large radii \mi\ and \mf\ stay relatively triaxial while the \mm\ CDM-only halo transitions to an oblate shape, which is unusual for a typical CDM-only simulated halo at this mass scale.  The DM+Baryon haloes \mf\ and \mi\ tend to be quite oblate at intermediate radii ($2\lesssim r \lesssim 50$ kpc), with very little difference between DM models.  At large radii ($r \gtrsim 50$ kpc) most haloes (even \mm) transition to a somewhat more triaxial shape.  This is also expected since the SIDM interaction rates (Figure \ref{fig:figr_Gscatter}) are quite low at these distances, so the DM behavior should not differ appreciably from CDM.  The degree of triaxiality at large radii varies substantially, however, with no clear trend with $\sigma/m$. 

In the inner regions of the haloes, \mf\ shows significant triaxiality in the CDM+Baryon case and more oblate structure in the SIDM+Baryon cases, while in \mm\ there is a huge variation in the degree of triaxiality in the central part of the halo.  Interestingly, the triaxiality of the \emph{total} mass distribution does not universally follow the stellar distribution in the inner galaxy the way that the $s=c/a$ axis ratio does.

\subsection{Comparison to Previous Work}

\begin{figure}
	\centering
	\includegraphics[width  = \columnwidth,
	                 height = \textheight,
	                 keepaspectratio]
	{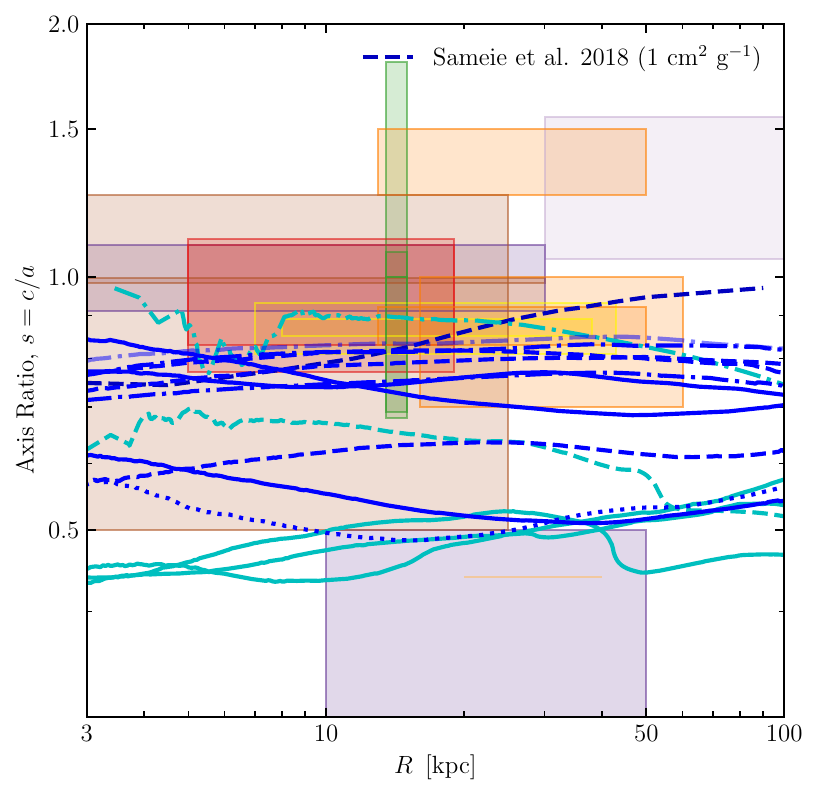}
	\caption{\textbf{Shape profiles of DM in simulated MW-mass galaxies compared to estimates for the MW.}  Axis ratio $s = c/a$ is shown as a function of geometrical mean radius $R = \left( abc \right)^{1/3}$ for the DM component in all simulations.  Line-styles and colours follow the legend in Figure \ref{fig:figr_legend} (but with thicker lines) for the simulations.  The dark blue dashed-line follows a previous SIDM+Baryon semi-analytic model for $\sigma/m=1$ cm$^2$ g$^{-1}$ \citep[see Figure 7 of][]{2018MNRAS.479..359S}.  The SIDM+Baryon simulations are generally more spherical than CDM-only, but not notably different than CDM+Baryon, and show less variation with $\sigma/m$ than in SIDM-only.  All simulations except CDM-only are loosely consistent with the various estimates for the MW.  For simulations with baryons, halo-to-halo variation is comparable to the difference between DM models.  The estimated values of the halo density flattening parameter $q_{\rho}^\mathrm{DM} = \left(c/a\right)_\rho$ are summarized in Table \ref{tbl:obs}, and are plotted over their approximate regions of validity (i.e. the range spanned by the data used for the constraints).  The coloured boxes and lines indicate the various estimates for $q_{\rho}^\mathrm{DM} = \left(c/a\right)_\rho$ summarized in Table \ref{tbl:obs}.  Constraints derived from models of individual tidal streams are shown in orange (Sagittarius stream), red (Palomar 5 stream), green (Grillmair-Dionatos-1 stream), and yellow (statistically detected streams around NGC3201 and M68).  Constraints based on Jeans modeling are shown in purple, and measurements combining equilibrium assumptions with the disc rotation curve and/or other disc data are shown in brown.}
	\label{fig:figr_ca_dark}
\end{figure}

We first compare the DM shape profiles to previous results from DM-only simulations, those with an analytic disc model, and CDM+Baryon simulations \citep[see][Figure 7 for a summary]{2018MNRAS.479..359S}.  In Figure \ref{fig:figr_ca_dark}, we plot the shape of the DM component for all the FIRE-2 MW-mass galaxy simulations of Table \ref{tbl:resims}.  This comparison uses log-scale for the $y$-axis axis ratio $s=c/a$ and geometric mean of the axis lengths $R=\left(abc\right)^{1/3}$ for the $x$-axis, for direct comparison with the summary in Figure 7 of \citet{2018MNRAS.479..359S}.  Plots of the axis ratio $s=c/a$ and triaxiality $T$ versus geometric mean radius $R$ for all individual species (as in Figures \ref{fig:figr_ca_all_sphericalr} and \ref{fig:figr_T_sphericalr}) are given in Appendix \ref{sec:geo}.

We see a clear trend towards more spherical haloes at larger $\sigma/m$ for the DM-only simulations, but find that the SIDM+Baryon simulations are not as spherical at larger radii as assumed in the SIDM+Baryon semi-analytic model of \citet{2018MNRAS.479..359S}, which is initialized with a spherically symmetric halo.  Also, instead of the concave-up shape predicted by this model for SIDM+Baryon (with the innermost and outermost regions the most spherical), we see a concave-down trend for all the curves (where the intermediate radii are most spherical).  This appears consistent with the idea that SIDM can respond more quickly to the influence of the central galaxy than CDM (which generally has a more spherical inner halo when compared across resimulations of the same initial conditions).  It is also consistent with the picture that CDM and SIDM should behave similarly in the halo outskirts, where the shape is driven mainly by the connection with the local filaments \citep[see e.g.][]{2011MNRAS.416.1377V} and therefore tends to be less spherical than at intermediate radii.  However, the variation in formation histories across the different haloes dominates over the variation with $\sigma/m$.

We also compare our results to estimates of the halo density flattening parameter (minor-to-major axis ratio) $q_{\rho}^\mathrm{DM} = (c/a)_\rho$ from the literature, summarized in Table \ref{tbl:obs}.  This quantity is sometimes referred to in the literature \citep[see e.g.][]{2020arXiv201203908H} as simply parameter $q$, but is changed here to distinguish from our intermediate-to-major axis ratio $q(r)=b(r)/a(r)$.  In modeling the kinematics of various MW tracers of the potential such as tidal streams, globular clusters, or ``field'' halo stars unassigned to a given stream, the parameter $q_{\rho}^\mathrm{DM}$ usually represents the flattening of the best-fitting axisymmetric NFW model for the DM density, and is comparable to our minor-to-major axis ratio $s(r)=c(r)/a(r)$.  In the case of \citet{2010ApJ...714..229L}, where a triaxial halo is used, we cite the value of $(c/a)_{\rho}$ quoted by the authors within 20 kpc, which was determined by fitting ellipsoids to the contours of the Laplacian of the potential.

\begin{table*}
    \caption{\textbf{Measurements of the shape of the MW halo from previous work.}  Uncertainties on $q_{\rho}^\mathrm{DM} = \left(c/a\right)_\rho$ parameter values are quoted as given in the various works and plotted in Figure \ref{fig:figr_ca_dark}, which usually correspond to 90\% confidence intervals or equivalent.  Values with no uncertainties have no easily interpretable range given in the corresponding paper, or are lower limits.  $r_{\mathrm{min}}$ and $r_{\mathrm{max}}$ denote the approximate range of galactocentric radii over which the estimates are made, either as given by the authors or as specified for the data-set used.  \emph{Notes}: (a) This paper includes the rotation curve, velocity dispersion, and vertical force profile of the disc as additional constraints.  (b) In these two cases, a flattening of $q_{\rho}^\mathrm{DM}=0.8$ in the density was assumed to detect the stream statistically before using it to fit a parametrized mass model.  (c) As pointed out by \citet{2020arXiv201203908H}, this paper finds a prolate aspect ratio using an action finder that has known difficulties for orbits in prolate mass distributions.  (d) As \citet{2010ApJ...714..229L} point out and \citet{2013MNRAS.434.2971D} confirm, this model does not admit a stable galactic disc; \citet{2013ApJ...773L...4V} shows that the discrepancy can be explained by the influence of the LMC.}
    \label{tbl:obs}
    \begin{tabular}{lrrrll}
    
        \hline
        \hline
        Reference                                      &
        $q_{\rho}^\mathrm{DM} = \left(c/a\right)_\rho$ &
        $r_{\mathrm{min}}$                             &
        $r_{\mathrm{max}}$                             &
        Data-set used                                  &
        Colour
        \\[1pt]
		                            &
		                            &
		$\left[ \text{kpc} \right]$ &
		$\left[ \text{kpc} \right]$ &
		                            &
		\\
		\hline
        
        \citet{2020arXiv201203908H} & $0.993^{+0.01}_{-0.005}$ &  1   &  30 & RR Lyrae + constraints\textsuperscript{a} & brown  \\
        \citet{2020arXiv201014381P} & $0.87^{+0.02}_{-0.02}$   &  8   &  38 & NGC3201\textsuperscript{b}                & yellow \\
        \citet{2019MNRAS.486.2995M} & $0.82^{+0.25}_{-0.13}$   & 13.5 &  15 & Grillmair-Dionatos-1 (GD-1)               & green  \\
        \citet{2019AA...621A..56P}  & $1.30^{+0.25}_{-0.25}$   & 30   & 150 & Globular clusters\textsuperscript{c}      & purple \\
        \citet{2019MNRAS.485.3296W} & $1.00^{+0.09}_{-0.09}$   &  1   &  30 & RR Lyrae                                  & purple \\
        \citet{2019MNRAS.488.1535P} & $0.87^{+0.06}_{-0.06}$   &  7   &  43 & M68\textsuperscript{b}                    & yellow \\
        \citet{2016ApJ...833...31B} & $1.3^{+0.5}_{-0.3}$      & 13.5 &  15 & Grillmair-Dionatos-1 (GD-1)               & green  \\
        \citet{2016ApJ...833...31B} & $0.93^{+0.16}_{-0.16}$   &  5   &  19 & Palomar 5 (Pal 5)                         & red    \\
        \citet{2015ApJ...803...80K} & $0.95^{+0.16}_{-0.12}$   &  5   &  19 & Palomar 5 (Pal 5)                         & red    \\
        \citet{2014ApJ...794..151L} & $0.4^{+0.1}_{-0.1}$      & 10   &  50 & SDSS halo stars                           & purple \\
        \citet{2010ApJ...712..260K} & $>0.68$                  & 13.5 &  15 & Grillmair-Dionatos-1 (GD-1)               & green  \\
        \citet{2010ApJ...714..229L} & $0.44$                   & 20   &  40 & Sagittarius (Sgr)\textsuperscript{d}      & orange \\
        \citet{2005ApJ...619..800J} & $0.88^{+0.04}_{-0.05}$   & 13   &  50 & Sagittarius (Sgr)                         & orange \\
        \citet{2004ApJ...610L..97H} & $1.30^{+0.20}_{-0.05}$   & 13   &  50 & Sagittarius (Sgr)                         & orange \\
        \citet{2001ApJ...551..294I} & $>0.7$                   & 16   &  60 & Sagittarius (Sgr)                         & orange \\
        \citet{2000MNRAS.311..361O} & $0.80^{+0.45}_{-0.30}$   &  1   &  25 & HI gas + disc rotation curve              & brown  \\
        \hline
        \hline
        
    \end{tabular}
\end{table*}

The various estimates for the shape of the MW (given by the references of Table \ref{tbl:obs}) vary as widely as shapes of the simulated galaxies, underlining the difficulty of the measurement.  These measurements have a wide spread in both $r$ and $q_{\rho}^\mathrm{DM}$, and thus, don't agree on the shape (or the triaxiality) of the MW.  One positive development from this work is that in most DM models the value of $s=c/a$ for the DM haloes of the simulated systems appears to be fairly constant ($\Delta s \lesssim 0.2$) over a wide range of radii ($3$--$100$ kpc), which should in principle simplify efforts to model the dark halo.  We caution, however, that (1) none of our models include a Large Magellanic Cloud (LMC)-like companion, which is likely to affect this assertion \citep[e.g.][]{2010ApJ...714..229L, 2013ApJ...773L...4V, 2020arXiv200910726V} and (2) that this statement assumes that any rotation of the principal axes with radius is precisely incorporated into the model.

\section{Discussion}
\label{sec:discussion}

The presence of baryons, and their resulting effect on the shapes of MW-mass galaxies, shows far wider variety than expected from DM-only and semi-analytic models.  Importantly, MW-mass galaxies in SIDM haloes at the preferred values of $\sigma/m$ based on studies of dwarf galaxies and galaxy clusters still have density, velocity, and shape profiles that are consistent with observations, as well as producing a realistic-looking disc galaxy at the centre.  Thus, there is no immediate discrepancy produced by the introduction of a nonzero self-interaction cross-section (at least in the velocity-independent, elastic collision model considered here) that can rule out this type of SIDM.

More interesting is the question of whether the variation in shape due to a nonzero cross-section could be constrained well enough to differentiate SIDM from CDM.  From Figure \ref{fig:figr_ca_dark} the hope of doing this seems fairly dim, since there is as much variation in shape from different assembly histories as from different DM models.  However, the importance of the question merits a closer look at comparisons involving the particular radii where we expect the differences to be largest.

The key region for looking at shape variations produced by SIDM is likely to be between about $2$--$20$ kpc, still inside $d_1$ (so the self-interactions have a chance to shape the system), but outside the region where the shape is utterly dominated by the central galaxy's baryons.  We have the advantage that in this region we can still tightly constrain the shape of the stellar and gas components from observations, and look instead at where the shape of the total mass distribution (constrained using dynamical modeling) departs from the shape of the stars.

\begin{figure*}
	\centering
	\includegraphics[width  = \textwidth,
	                 height = \textheight,
	                 keepaspectratio]
	{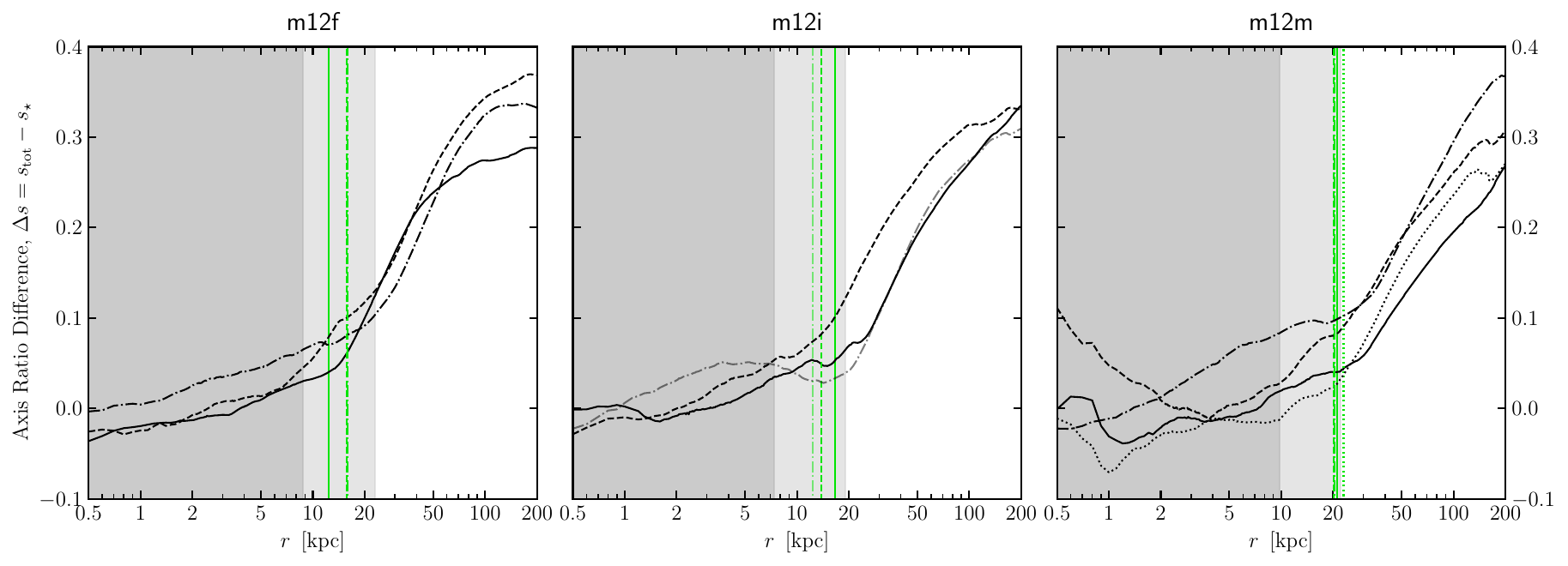}
	\caption{\textbf{Difference in shape profiles of simulated MW-mass galaxies.}  Minor-to-major axis ratio difference $\Delta s = s_\mathrm{tot}-s_\star$ is shown as a function of semi-major axis $r=a$ in three sets of MW-mass galaxy simulations: \mf\ (left), \mi\ (centre), and \mm\ (right).  Line-styles and shaded-areas follow the legend in Figure \ref{fig:figr_legend}.  The black curves represent the axis ratio difference $\Delta s$ and the vertical lines represent $r_{\star,90}$, the spherical radius of 90\% of stellar mass within 30 kpc of the central galaxy, given in Table \ref{tbl:resims}.}
	\label{fig:figr_Delta_s_vs_r}
\end{figure*}

We do this by plotting the difference between the total and stellar axis ratios $\Delta s = s_\mathrm{tot}-s_\star$ in Figure \ref{fig:figr_Delta_s_vs_r}.  The region $r<5$ kpc is dominated by the bulge dynamics and hence varies substantially between galaxies.  However, outside this region we see that there are clear transitions in the slope for all the curves between the flattened, stellar-disc-dominated regime (out to around 20 kpc) and the region where the total shape is determined by the more spherical DM halo.  For CDM+Baryon simulations, the transitions roughly occur at $r_{\star,90}$, represented by the green vertical lines.  For the SIDM+Baryon simulations, the transitions roughly occur at $d_1$, represented by the shaded-areas, following the legend in Figure \ref{fig:figr_legend}.  While the CDM+Baryon transitions are largely dependent on the stellar mass, the SIDM+Baryon transitions are instead dependent on the DM self-interaction cross-section.  This holds for $\sigma/m=1$ and 10 cm$^2$ g$^{-1}$, but not for the \mm\ SIDM+Baryon $\sigma/m=0.1$ cm$^2$ g$^{-1}$, which has no detectable $d_1$ at $z=0$.  Since the density for $\sigma/m=0.1$ cm$^2$ g$^{-1}$ does not reach the levels needed for local collisions to occur within the age of the halo, this cross-section behaves similarly to CDM and the shape transition occurs at $r_{\star,90}$.  Likewise, constructing such a test would be difficult near $\sigma/m=10$ cm$^2$ g$^{-1}$ in \mm\ since $d_1$ and $r_{\star,90}$ are very close in this case.  Otherwise, we see that for SIDM+Baryon, across all three simulations, \emph{the shape of the total mass distribution departs from that of the stellar mass distribution at steadily increasing radius as $\sigma/m$ increases}.  Thus, for a given galaxy, constraining the radius of the transition from where stars dominate its shape to where DM is the dominant influence and comparing this to (1) the galaxy scale length and (2) the predicted $d_1$ as a function of $\sigma/m$ provides a way to constrain the SIDM cross-section.

Currently, there are several methods in use for constraining the shape of the MW's total mass distribution using dynamical modeling.  A non-exhaustive sample of measurements using these different methods is listed in Table \ref{tbl:obs}.  One strategy is to model the precession of the orbital planes of tidal streams.  Several attempts to do this for the Sagittarius stream alone \citep[][]{2004ApJ...610L..97H, 2005ApJ...619..800J, 2009ApJ...703L..67L, 2013ApJ...773L...4V, 2020arXiv200910726V} have come to inconsistent conclusions, mostly due to differing parametrizations of the potential and the treatment of the influence of the LMC on the outer portion of the stream (as pointed out in \citealt{2013ApJ...773L...4V} and \citealt{2020MNRAS.498.5574E}).  However, since we need mainly to constrain the region $2\lesssim r \lesssim 20$ kpc for these tests, simultaneous fits of several streams with smaller apocenters may provide a way forward \citep[][]{2016ApJ...833...31B, 2018ApJ...867..101B, 2020arXiv200700356R}.

Another strategy is to constrain the total flattening using Jeans or Schwarzschild modeling of equilibrium stellar populations as in \citet{2014ApJ...794..151L, 2020arXiv201203908H}, and many other works.  However, this is limited to regions where we have sufficient stellar tracers observed to derive the distribution functions used in the model.  Currently such efforts have been made in the bulge (to $r\sim 5$ kpc) and in the space observatory Gaia satellite's six-dimensional (6D) volume ($5 \lesssim r \lesssim 11$ kpc).  However, additional data from ground-based spectroscopic surveys, future Gaia data releases, and new distance estimators \citep[e.g][]{2019MNRAS.484..294D,2020AJ....160...18A} promise to expand the volume accessible to this technique appreciably in the near future.  These new data will also provide excellent constraints on the \emph{stellar} shape profile, an equally important quantity in this approach.

A third strategy, as employed by e.g. \citet{1991ApJ...370..205B, 2000MNRAS.311..361O, 2005A&A...440..523N} in the MW and \citet{2020ApJ...889...10D} for external galaxies, is to use the flaring of the HI disc to constrain the flattening, under the assumption that the gas is in dynamical equilibrium.  Results seem to favor a relatively spherical halo with $q_{\rho}^\mathrm{DM} \sim 0.8$.  This technique could provide an independent assessment with different data and systematics than methods using stellar kinematics. 

Finally, a lower limit on $s=c/a$ as a function of radius may be obtained by searching for evidence of the truncation or scattering of tidal streams by orbital resonances, which are more common in more highly flattened potentials \citep[][]{2012MNRAS.419.1951V, 2018arXiv180403670H, 2015ApJ...799...28P, 2020arXiv200909004V}.  These effects are quite pronounced at $c/a \sim 0.3$--$0.4$, where most of our shape profiles begin their transition toward the more spherical halo, and are much less dominant by $c/a \sim 0.6$--$0.7$, where most of the total mass profiles end up at large $r$.  Looking for an abrupt transition in the prevalence of streams as a function of galactocentric distance, perhaps even in a stacked sample of external galaxies where 6D information is not available, could be an additional way to find constraints on the halo flattening transition and hence on SIDM.

\section{Conclusion}
\label{sec:conclusion}

We perform a suite of cosmological-baryonic zoom simulations of Milky Way (MW)-mass galaxies for several different models with self-interacting dark matter (SIDM), one proposed solution to the challenges of the cold dark matter (CDM) plus dark energy $\left( \Lambda \text{CDM} \right)$ cosmological model at small-scales.  These dark matter (DM) simulations are compared between CDM and SIDM (with interaction cross-sections $\sigma/m=0.1$, 1, and 10 cm$^2$ g$^{-1}$) and with expectations from the literature.

For the SIDM+Baryon simulations, the variation in axis ratio with SIDM cross-section is not as large as expected in the literature.  The assembly history of the central galaxy is the dominant influence inside the local collision region (LCR).  Variations in the assembly and evolution of the galaxy thus dominate the resulting shape.  At larger radii, greater differences between CDM and SIDM axis ratios are also expected according to DM-only models, but again the halo-to-halo variation in the assembly history of DM+Baryon causes larger differences in halo shape than the variation in $\sigma/m$ does.  In general, the flattening profile with radius tends to be concave-down (most spherical at intermediate radii) rather than concave-up (most spherical at small and large radii) as predicted by previous work.

Although we find that halo-to-halo variation is larger than variation due to different $\sigma/m$, a possible test for SIDM could lie in predictions of the difference between the shape of the stellar distribution and that of the total mass distribution, both of which can be constrained by different methods.  The overall shape of the total distribution (inferred from dynamics) is dominated by the flattened stellar component inside the LCR and by the more spherical DM component outside the LCR.  The radius of this transition occurs at radius of 90\% of stellar mass $r_{\star,90}$ for CDM+Baryon simulations, but increases with increasing $\sigma/m$ and occurs at the LCR scattering radius $d_1$ for SIDM+Baryon simulations, as the response of the DM to the growing galaxy becomes more important.  There are several promising possibilities for measuring this radius with new survey data in the MW, and perhaps in other galaxies, in the coming decade. 

Our results at very small radius ($\lesssim 2$ kpc) are limited by the need for better DM particle resolution for the SIDM-only simulations particularly, since as predicted the self-interactions significantly heat the central halo and reduce the particle density, limiting our ability to measure the shape in this region with our Lagrangian approach to simulations.  In the case of all DM+Baryon simulations, this limitation is mitigated by the presence of the central galaxy, which deepens the gravitational potential and boosts the DM density in nearly all cases.  However, better resolution in the central part would enable us to better study the transport of energy and angular momentum between the stellar and DM components, which will be the subject of future work.

Finally, we note that the galaxies formed in SIDM haloes with baryonic feedback differ from those in CDM in mostly subtle ways, and are generally similar (and consistent with observations) in their large-scale properties such as mass and scale radius.  Therefore, SIDM remains a valid possibility for new DM physics.

\section*{Acknowledgements}

RES acknowledges support from NASA grant 19-ATP19-0068 and HST-AR-15809 from the Space Telescope Science Institute (STScI), which is operated by AURA, Inc., under NASA contract NAS5-26555.  MBK acknowledges support from NSF CAREER award AST-1752913, NSF grant AST-1910346, NASA grant NNX17AG29G, and HST-AR-15006, HST-AR-15809, HST-GO-15658, HST-GO-15901, and HST-GO-15902 from STScI.  AW received support from NASA through ATP grants 80NSSC18K1097 and 80NSSC20K0513; HST grants GO-14734, AR-15057, AR-15809, and GO-15902 from STScI; a Scialog Award from the Heising-Simons Foundation; and a Hellman Fellowship.  ASG is supported by the Harlan J. Smith postdoctoral fellowship.  This research is part of the Frontera computing project at the Texas Advanced Computing Center (TACC).  Frontera is made possible by National Science Foundation award OAC-1818253.  Simulations in this project were run using Early Science Allocation 1923870, and analysed using computing resources supported by the Scientific Computing Core at the Flatiron Institute.  This work used additional computational resources of the University of Texas at Austin and TACC, the NASA Advanced Supercomputing (NAS) Division and the NASA Center for Climate Simulation (NCCS), and the Extreme Science and Engineering Discovery Environment (XSEDE), which is supported by National Science Foundation grant number OCI-1053575.

\section*{Data Availability}
The simulations used for this study are currently proprietary to members of the FIRE collaboration. Please contact the authors if interested.
 



\bibliographystyle{mnras}
\bibliography{mnras_SIDM_Structure} 




\appendix

\section{Ellipsoidal Versus Spherical Shells} \label{sec:evs}

\begin{figure}
	\centering
	\includegraphics[width  = \columnwidth,
	                 height = \textheight,
	                 keepaspectratio]
	{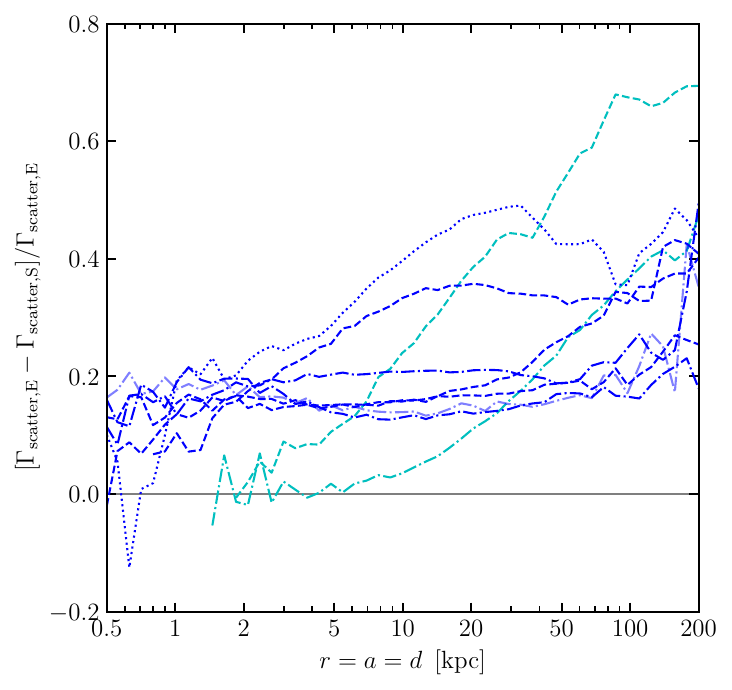}
	\caption{{\bf Difference in DM local collision rate of simulated MW-mass galaxies.} The difference when averaging over shells of ellipsoidal distances $d$ versus spherical radii $r=a$, $\left[ \Gamma_\mathrm{scatter,E}(d)-\Gamma_\mathrm{scatter,S}(r) \right] /\Gamma_\mathrm{scatter,E}(d)$ (Equation \ref{eqn:R_scatter}) as a function of $d=r=a$ for SIDM simulations.  Shells are matched so that spherical radius $r$ is equal to the semi-major axis $a$ of the ellipsoid used to compute $d$.  Line-styles and colours follow the legend in Figure \ref{fig:figr_legend}.  Using ellipsoidal shells leads to larger $\Gamma_\mathrm{scatter}$, mostly because of an increase in the computed DM density.  The difference in the local collision rate increases for the ellipsoidal shells method away from the centre of the galaxies.}
	\label{fig:figr_GscatterES}
\end{figure}

As discussed in \S \ref{sec:results}, we use the triaxial ellipsoidal shape profiles to calculate density and velocity profiles in ellipsoidal shells.  However, past work has largely relied on spherical shells to calculate these profiles.  Figure \ref{fig:figr_GscatterES} shows the difference in DM local collision rate $\left[ \Gamma_\mathrm{scatter,E}(d)-\Gamma_\mathrm{scatter,S}(r) \right] /\Gamma_\mathrm{scatter,E}(d)$ between ellipsoidal shells and spherical shells, where $\Gamma_\mathrm{scatter,E}(d)$ and $\Gamma_\mathrm{scatter,S}(r)$ are the $\Gamma_\mathrm{scatter}$ for ellipsoidal and spherical methods, respectively.  We note that the ellipsoidal shells have larger $\Gamma_\mathrm{scatter}$ due to having larger estimates of the DM density, particularly at larger $d$.  This is the effect of ``smearing'' across isodensity contours when using spherical shells to compute densities.  The DM velocity profiles have negligible differences between the ellipsoidal and spherical methods, suggesting that the local velocity ellipsoid is relatively isotropic.

We calculate the difference in scattering radius when using ellipsoidal shells ($d_1$) rather than spherical shells ($r_1$) as $\Delta_1=d_1-r_1$, where $\Delta_1=2.$ kpc for SIDM-only $\sigma/m = 10$ cm$^2$ g$^{-1}$.  For SIDM+Baryon $\sigma/m = 1$ and 10 cm$^2$ g$^{-1}$, the differences are $\Delta_1=0.6$--$1.6$ kpc and $\Delta_1=1.$--$2.$ kpc, respectively.  The $\Delta_1 \sim 5$--$20$\% difference demonstrates the importance of using ellipsoidal shells fit to the isodensity contours to estimate density and velocity profiles.  The use of ellipsoidal shells is even more important for the stars, gas, and total mass distributions, since these have less spherical shapes, and thus larger density and velocity differences between the two methods compared to the DM component.

\section{Geometrical mean of axis lengths} \label{sec:geo}

\begin{figure*}
	\centering
	\includegraphics[width  = \textwidth,
                 	 height = \textheight,
	                 keepaspectratio]
	{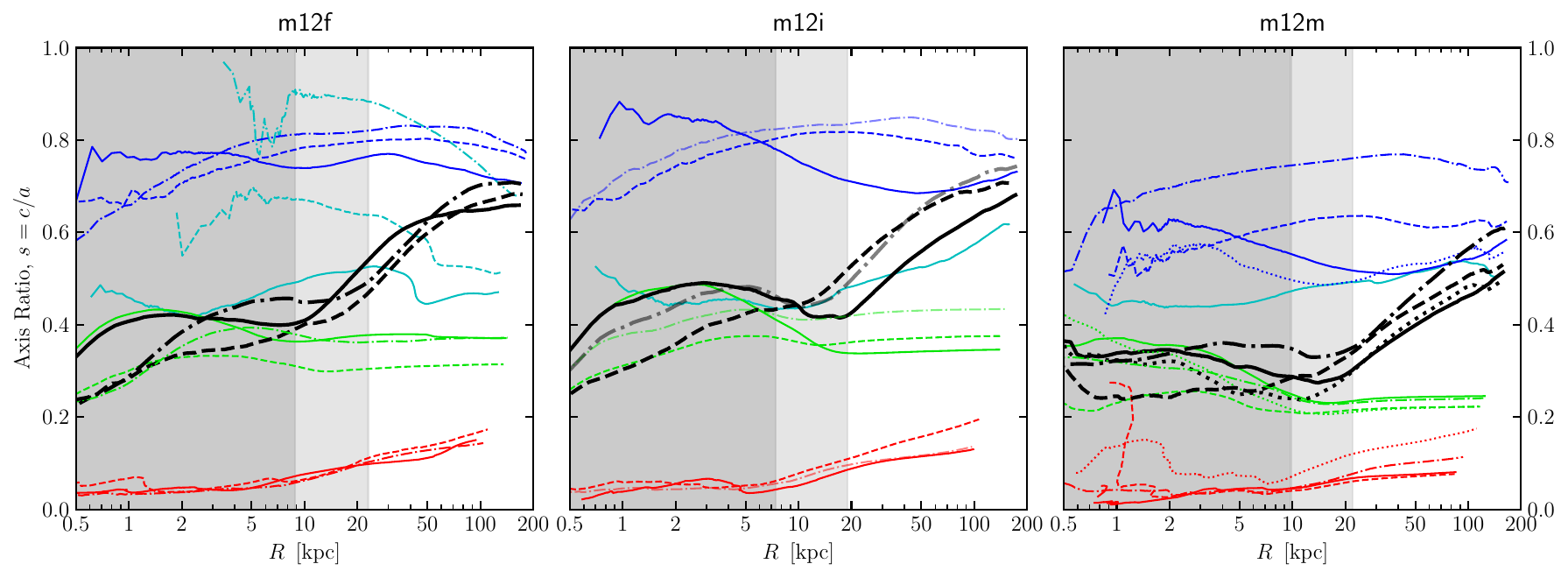}
	\caption{\textbf{Shape profiles of simulated MW-mass galaxies as a function of $R$.}  Minor-to-major axis ratio $s=c/a$ as a function of geometrical mean radius $R$ for different species (DM, stars, and gas) in three sets of MW-mass galaxy simulations: \mf\ (left), \mi\ (centre), and \mm\ (right).  Line-styles, colours, and shaded-areas follow the legend in Figure \ref{fig:figr_legend}.  An alternate version of this plot using the semi-major axis $r=a$ can be found in the main body of the work in Figure \ref{fig:figr_ca_all_sphericalr}.  The literature often plots axis ratios versus this geometrical mean of the axis lengths $R$, which is therefore given here for comparison purposes.  We note that the main difference is in how the curves have shifted leftward towards smaller radii.}
	\label{fig:figr_ca_all_geometricR}
\end{figure*}

Figure \ref{fig:figr_ca_all_geometricR} shows the axis ratio $s=c/a$ versus the geometrical mean of the axis ratios $R=\left(abc\right)^{1/3}$, instead of semi-major axis distance $r=a$ as in Figure \ref{fig:figr_ca_all_sphericalr}.  This plot is included to facilitate comparison with previous work on halo shapes.  The most important difference to note relative to Figure \ref{fig:figr_ca_all_sphericalr} is that the axis ratio curves have shifted towards smaller radii, since the ellipsoidal geometrical mean radius $R$ is always less than or equal to semi-major axis $r=a$ by definition.  Due to the relatively spherical shape of the DM component, its curve is shifted the least, while the profile of the gas component, which is the most flattened, has shifted the most.  Rapid changes in the flattening of neighboring ellipsoids have the effect of producing non-functional curves, such as observed for the gas in \mm\ SIDM+Baryon $\sigma/m = 1$ cm$^2$ g$^{-1}$ near $R \sim 1$ kpc.  This is a result of the fact that the definition of $R$ does not guarantee that it must always increase with increasing ellipsoid semi-major axis $r=a$.

\begin{figure*}
	\centering
	\includegraphics[width  = \textwidth,
                 	 height = \textheight,
	                 keepaspectratio]
	{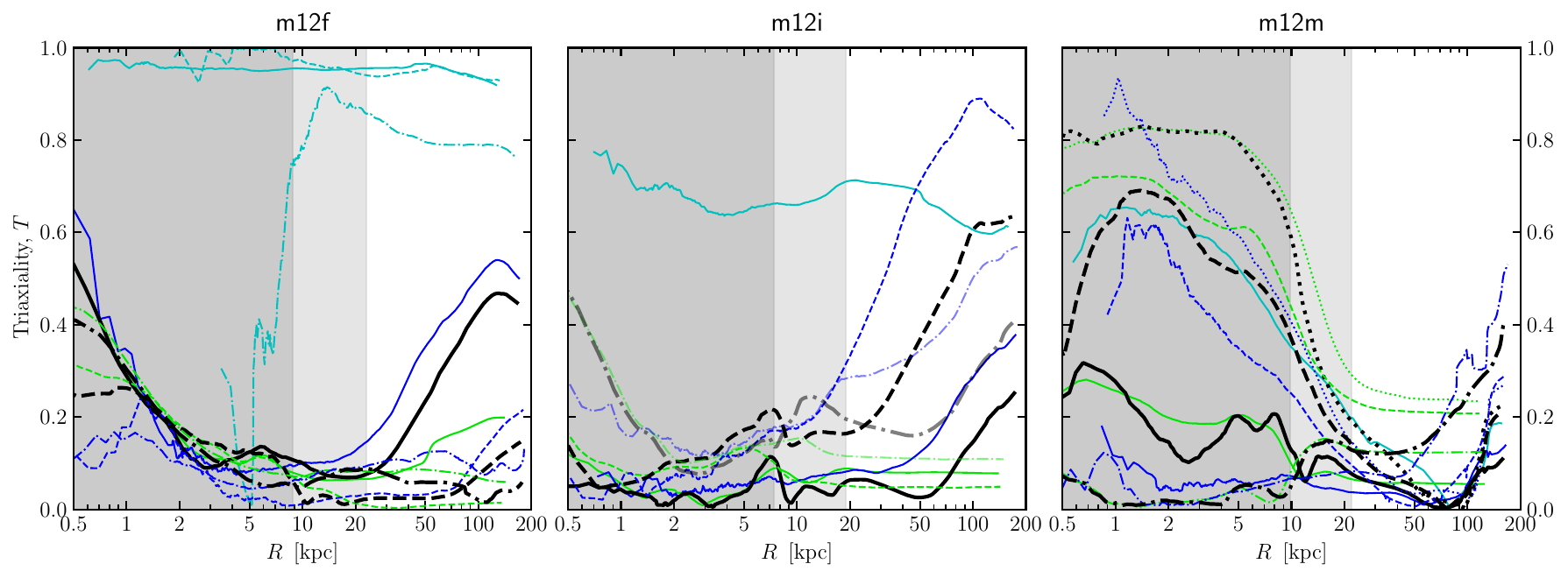}
	\caption{\textbf{Triaxiality profiles of simulated MW-mass galaxies as a function of $R$.}  Triaxiality $T$ as a function of geometrical mean radius $R$ for different species (DM and stars) in three sets of MW-mass galaxy simulations: \mf\ (left), \mi\ (centre), and \mm\ (right).  Line-styles, colours, and shaded-areas follow the legend in Figure \ref{fig:figr_legend}.  An alternate version of this plot using the semi-major axis $r=a$ can be found in the main body of the work in Figure \ref{fig:figr_T_sphericalr}.  The literature often plots triaxiality versus this geometrical mean of the axis lengths $R$, which is therefore given here for comparison purposes.  We note that the main difference is in how the curves have shifted leftward towards smaller radii.}
	\label{fig:figr_T_geometricR}
\end{figure*}

Similarly, Figure \ref{fig:figr_T_geometricR} shows the triaxiality $T$ versus the geometrical mean of the axis lengths $R$, instead of semi-major axis $r=a$.  The plot shows a similar shift leftward for all the triaxiality curves, but the shift is less pronounced since the triaxiality depends on all three axis lengths $a$, $b$, and $c$.  This shift does in general reduce the apparent prolateness of the CDM+Baryon and SIDM+Baryon MW-mass galaxies at small radii, especially for \mf\ and \mi.

Overall, we consider that the use of semi-major axis length $r=a$ is preferable to the use of the geometrical mean of the axis lengths $R$, since observations of galaxies are 3D light and velocity (redshift) distributions projected as two-dimensional (2D) light and velocity distributions on to the celestial sphere.  Therefore, measuring a realistic $R$ for any galaxy is a difficult and degenerate task.  On the other hand, deprojecting \emph{only} the semi-major and semi-minor axis-lengths of nested ellipsoids is a more feasible task, since measurements of the line-of-sight velocity distribution \citep[e.g.][]{2018ApJ...863L..19L, 2020MNRAS.491.1690J} or stacked observations \citep[e.g.][]{2020ApJ...900..163K} can be used to constrain or marginalize over the inclination angle.  This motivates the use of $r=a$, inside which density, velocity, and shape profiles of the baryons (and thus also DM) are then more easily estimated.


\bsp	
\label{lastpage}
\end{document}